\documentclass[a4paper,fleqn,usenatbib]{mnras}
\usepackage{mathptmx}

\usepackage[T1]{fontenc}
\usepackage{ae,aecompl}
\usepackage[utf8]{inputenc}
\usepackage[czech]{babel} 
\usepackage{lmodern}
\usepackage{multirow}
\usepackage{array}

\usepackage{graphicx}	
\usepackage{amsmath}	
\usepackage{amssymb}	
\usepackage[flushleft]{threeparttable}

\newcommand{\Msun}{{\mathrm M}_{\odot}}

\newcommand{\nH}{n_{\mathrm{H}}}

\newcommand{\ccm}{\,\mathrm{cm}^{-3}}
\newcommand{\Zsol}{{\mathrm Z}_{\odot} }
\newcommand{\kms}{\,\mathrm{km\,s}^{-1}}
\newcommand{\K}{\,\mathrm{K}}

\defcitealias{haardt_radiative_2012}{HM12}

\title[Radiative feedback in faint $z>6$ galaxies]{Does radiative feedback make faint $z>6$ galaxies look small?}

\author[S. Ploeckinger et al.]{
Sylvia~Ploeckinger,$^{1,2}$\thanks{E-mail: ploeckinger@strw.leidenuniv.nl}
Joop~Schaye,$^1$
Alvaro~Hacar,$^1$ 
Michael~V.~Maseda,$^1$ 
\newauthor
Jacqueline~A.~Hodge,$^1$
Rychard~J.~Bouwens$^1$
\\
$^{1}$Leiden Observatory, Leiden University, PO Box 9513, NL-2300 RA Leiden, the Netherlands\\
$^{2}$Institute for Computational Cosmology, Durham University, South Road, Durham DH1 3LE, UK\\
}

\date{Accepted XXX. Received YYY; in original form ZZZ}

\pubyear{2018}

\begin{document}
\label{firstpage}
\pagerange{\pageref{firstpage}--\pageref{lastpage}}
\maketitle

\begin{abstract}

Recent observations of lensed sources have shown that the faintest ($M_{\mathrm{UV}} \approx -15\,\mathrm{mag}$) galaxies observed at $z=6-8$ appear to be extremely compact. Some of them have inferred sizes of less than 40 pc for stellar masses between $10^6$ and $10^7\,\Msun$, comparable to individual super star clusters or star cluster complexes at low redshift. High-redshift, low-mass galaxies are expected to show a clumpy, irregular morphology and if star clusters form in each of these well-separated clumps, the observed galaxy size would be much larger than the size of an individual star forming region. 
As supernova explosions impact the galaxy with a minimum delay time that exceeds the time required to form a massive star cluster, other processes are required to explain the absence of additional massive star forming regions. In this work we investigate whether the radiation of a young massive star cluster can suppress the formation of other detectable clusters within the same galaxy already before supernova feedback can affect the galaxy. 
We find that in low-mass ($M_{200} \lesssim 10^{10}\,\Msun$) haloes, the radiation from a compact star forming region with an initial mass of $10^{7}\,\Msun$ can keep gas clumps with Jeans masses larger than $\approx 10^{7}\,\Msun$ warm and ionized throughout the galaxy. 
In this picture, the small intrinsic sizes measured in the faintest $z=6-8$ galaxies are a natural consequence of the strong radiation field that stabilises massive gas clumps. A prediction of this mechanism is that the escape fraction for ionizing radiation is high for the extremely compact, high-$z$ sources.

\end{abstract}




\section{Introduction}\label{sec:intro}

Gravitational lensing of galaxy clusters permits resolved studies of star-forming regions of high-redshift galaxies beyond the detection limit of current telescopes. The combination of long exposure times with state-of-the art instruments such as the Wide Field Camera 3 (WFC3) on the Hubble Space Telescope (HST) and detailed lensing models allows one to reconstruct the original, un-lensed image of the background galaxy and to derive its physical properties. Recently, compact star-forming regions with spatial extents of less than 100 proper pc have been identified at redshifts higher than 6 \citep{bouwens_very_2017, vanzella_paving_2017} in the Hubble Frontier Fields \citep[HFF,][]{coe_frontier_2015, lotz_frontier_2017}. \citet{bouwens_very_2017} focus on the first four HFF clusters with the most refined lensing model and due to the depth of the HFF observations and the lensing from the massive foreground clusters, they claim to measure sizes to a typical $1 \sigma$ accuracy of $\approx$ 10 pc at $M_{\mathrm{UV}} \approx -15\,\mathrm{mag}$ and $\approx$ 50 pc at $M_{\mathrm{UV}} \approx -18\,\mathrm{mag}$.
The most compact sources in their sample are candidate young globular clusters (GCs) as both their sizes (< 40 pc) as well as their luminosities and derived stellar masses are similar to those of single star cluster complexes in the local Universe \citep[see e.g. fig. 9 in][]{bouwens_very_2017}.

The formation mechanism of GCs is still under debate \citep[see][for a recent review on GCs]{forbes_globular_2018}. 
\citet{kruijssen_globular_2015} proposed that high gas pressure leads to the formation of massive clusters, which are observed at low redshift as young, massive clusters (YMCs) or as old globular clusters. In their model, low-redshift disc galaxies do not reach the required high gas pressures in the absence of perturbations such as major mergers. On the other hand, in turbulent, gas rich discs, as are typical for high-redshift galaxies, GCs can form directly in the disc. If an unstable disc fragments and produces gas clumps with masses higher than the Jeans mass \citep{kim_amplification_2001}, the individual clumps can collapse and form star clusters massive enough for GC progenitors. 

If the population of very compact star-forming systems found in the HFF correspond to forming GCs-like objects (or a complex of forming GCs),
the measured extremely small sizes would indicate that one compact star cluster complex dominates the faintest ($M_{\mathrm{UV}} \approx -15\,\mathrm{mag}$) galaxies at that time. Why would only one region of $<$ 40 pc within the galaxy form a compact star cluster that is bright enough to be observed? 
In the sample of \citet{bouwens_very_2017} 46 (83) per cent of galaxies with $M_{\mathrm{UV}} > -16$\footnote{A UV magnitude of $M_{\mathrm{UV}} = -16$ corresponds to a stellar mass $\log M_{\star} [\Msun] \approx 6.5 - 7$ \citep[see figures 9 and 10 in][]{bouwens_very_2017}. } have sizes below 40 (100) pc. The measured size-luminosity relations for these faint objects in the samples of \citet{laporte_young_2016} and \citet{kawamata_sizeluminosity_2018} follow a similar trend. 

These sizes are significantly smaller than expected from the extrapolated size-luminosity relation of the brighter galaxies in the sample. Following this extrapolated relation, the sources would have median sizes of 100 pc at $M_{\mathrm{UV}} = -15$, compared to the inferred median sizes of $\approx$ 40 pc. Assuming that the compact sources from \citet{bouwens_very_2017} are embedded in a larger but lower surface brightness galaxy, the strong stellar feedback from the observed YMC (complex) could prevent star formation in the other gas clumps in the galaxy. This would result in the observed small sizes of the stellar component if only the bright star cluster is above the detection limit, as seen e.g. in \citet{ma_simulating_2018} for size measurements of simulated high-redshift ($z \ge 5$) galaxies for different surface brightness limits. 

The effect of supernova (SN) feedback has been extensively studied on various scales, for example, in idealised simulations of the interstellar medium (ISM; e.g. SILCC: \citealp{walch_silcc_2015, girichidis_silcc_2016}, TIGRESS: \citealp{kim_three-phase_2017}), dwarf galaxies in cosmological zoom-in simulations (e.g. FIRE: \citealp{hopkins_galaxies_2014, muratov_gusty_2015}), and cosmological volumes (e.g. EAGLE: \citealp{schaye_eagle_2015}, Illustris: \citealp{vogelsberger_introducing_2014}). While SNe can drive powerful winds, which can destroy dense gas clouds and quench star formation, there is a delay between the formation of the star cluster and the time when the SN wind has propagated through the galaxy. This delay time $\tau_{\mathrm{SN}}$ between the formation of the YMC and the galaxy-wide quenching of star formation by SN winds can be estimated from the lifetime of a $100\,\Msun$ star ($\approx 3.3\,\mathrm{Myr}$, \citealp{portinari_galactic_1998, leitherer_effects_2014}) plus the time the SN wind needs to cross a galaxy with size $R_{\mathrm{eff}}$, which depends on the wind velocity $v_w$:

\begin{equation} \label{eq:SNtimescale}
	\tau_{\mathrm{SN}} = 3.3\,\mathrm{Myr} + 19.6 \,\mathrm{Myr}\,\left ( \frac{R_{\mathrm{eff}}}{\mathrm{kpc}}  \right ) \left ( \frac{v_w}{50\,\kms} \right )^{-1}  ,
\end{equation}

\noindent using a reference wind velocity of $50\,\kms$ in line with the argument made in \citet{bouwens_very_2017}. 
 
 The formation time of high-redshift YMCs is not well constrained. Observations of local YMCs reveal small spreads in the ages of the member stars, e.g. 0.4 Myr for the most massive YMC known in the Galaxy (Westerlund 1, \citealp{kudryavtseva_instantaneous_2012}). In addition, other local YMCs with ages of a few Myr, such as the Arches (2.5-4 Myr), NGC 3603 (1-2 Myr), Trumpler 14 ($\approx$1 Myr) have already cleared out their dense gas \citep[for a review see e.g.][]{longmore_formation_2014}. The small age spread, as well as the short time-scale for gas removal indicate that star clusters form faster than the SN delay time $\tau_{\mathrm{SN}}$. Therefore, additional star clusters could form within a galaxy, unaffected by SN feedback. 

If a perturbation in the host galaxy leads to (disc) fragmentation that produces the observed bright sources at high redshift, it could be expected that other similarly massive clusters form before SNe have had the opportunity to drive the gas out of the disc. However, additional bright star clusters in the same galaxy would increase the galaxy sizes noticeably, which would be in tension with observations of the faint high-$z$ galaxies. \citet{boylan-kolchin_little_2018} suggested that globular clusters contribute considerably to the high redshift UV luminosity function, which would indicate that a large fraction of stars form in massive, dense star clusters in high redshift galaxies. 
This different mode of star formation (in contrast to less clustered star formation in local galaxies) is likely caused by the high gas surface densities \citep[$> 500-1000\,\Msun\,\mathrm{pc}^{-2}$, see e.g.][]{elmegreen_two_2018}, where the higher disc pressures can create more compact star forming regions. In this work, we test the impact of the strong radiation field from these compact sources on the total star formation in the host galaxy. If one YMC suppresses low-mass star formation in the remaining parts of the galaxy, it could be an explanation for the possible lack of unclustered star formation at high redshifts.

The impact of the strong radiation from young star clusters has been studied as an intrinsic feedback process to stop the collapse of the birth cloud \citep[e.g.][]{krumholz_dynamics_2009, abe_suppression_2018} as well as to drive galaxy-wide feedback: 
In the high-resolution, radiation hydrodynamic simulations of disc galaxies by \citet{rosdahl_galaxies_2015}, the radiation from young stars efficiently suppresses star formation in the disk, though it does not quench it. Different from SN feedback, which can destroy massive gas clumps that have already formed, radiation feedback predominantly prevents the formation of clumps, mainly by photoionization heating. This leads to a smoother density field and therefore less clustered star formation. Recently, \cite{guillard_impact_2018} confirmed this result by following the formation and destruction of individual star clusters in an isolated galaxy. They also find a significantly reduced number of star clusters compared to a simulation with identical initial conditions but only SN feedback. While these simulations illustrate the effect of radiative processes for individual isolated disc galaxies, they do not directly provide a generalised theoretical framework.

Here, we explore analytically whether radiative feedback from YMCs such as those observed by  \citet{bouwens_very_2017} can suppress further massive star cluster formation in high-$z$ galaxies prior to the first SNe. 
We use a simple analytic model as well as numerical radiative transfer calculations to identify how an individual massive star cluster to prevent the formation of other massive star clusters in a high-redshift dwarf galaxy by keeping gas clumps warm and ionized that could otherwise cool and form stars. For galaxies small enough to have all their gas ionized by the YMC, the observed sizes of the galaxy correspond to the size of the YMC. For larger or more massive galaxies where additional gas clumps can collapse and form new YMCs, the observed size is more representative of the gas disc of the galaxy.  
(Radiative) feedback can therefore explain both the small observed sizes in the faint $z=6$ galaxies as well as the inferred change in the slope of their luminosity-size relation. 

The general idea as well as the key results are summarised in Fig.~\ref{fig:sketchradfield}. 
The analytic models used to determine if one YMC can suppress star formation galaxy-wide by radiative feedback are described in Sec.~\ref{sec:methods} for a gas distribution with constant density (Sec.~\ref{sec:homogeneous}) and for a clumpy gas structure (Sec.~\ref{sec:clumpy}). Numerical results from the spectral synthesis code \textsc{Cloudy} \citep[version 17.00,][]{ferland_cloudy_1998, ferland_2013_2013, ferland_2017_2017} in Sec.~\ref{sec:results} provide additional information on the thermal state of the gas as well as the sensitivity of the results to the metal and dust content of the gas. The implications of this study for the escape fraction of ionizing photons are discussed in Sec.~\ref{sec:discussion}. We summarise our findings in Sec.~\ref{sec:summary}.

\begin{figure}
	\begin{center}
		\includegraphics[width=\linewidth, trim = 3.5cm 8.cm 3.5cm 5.5cm, clip]{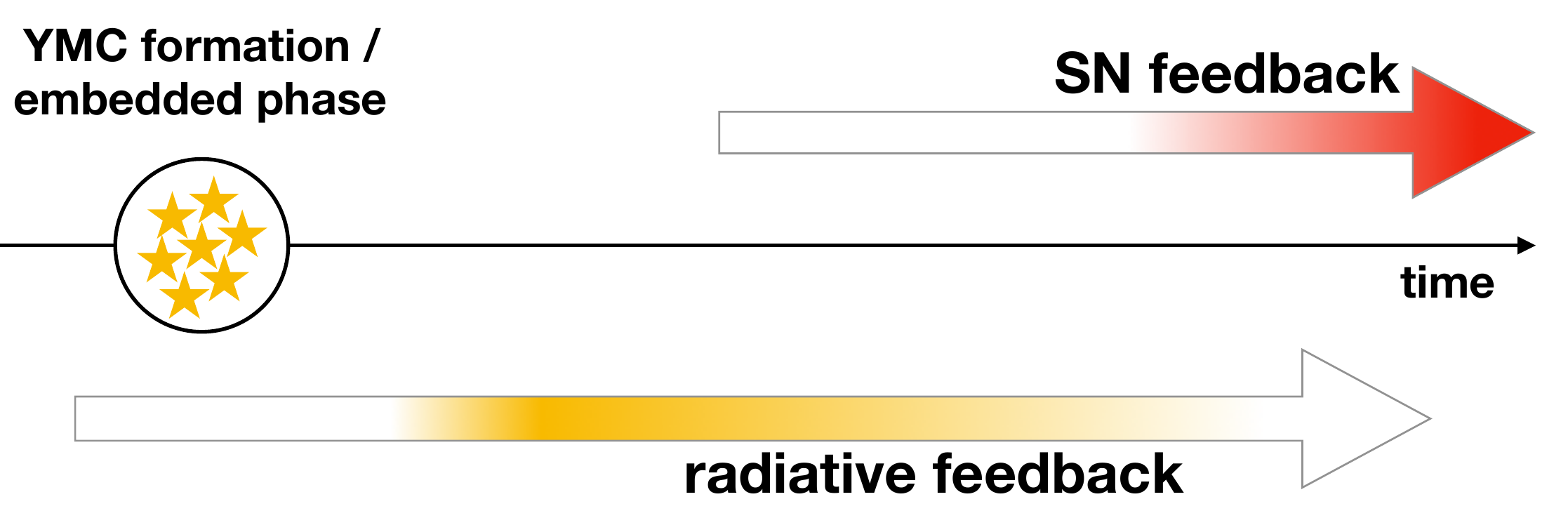}
		\caption{Timeline for the different feedback processes. Ionizing radiation from the YMC can impact the rest of the galaxy as soon as the YMC clears out its birth cloud and leaves the embedded phase. It is most effective during the first few Myr after the embedded phase. The ionizing luminosity of the YMC decreases rapidly with age after the first stars in the cluster reach the end of their lifetimes (see Table~\ref{tab:overview}). The impact of SN feedback increases as the SN-driven wind propagates through the galaxy. }
		\label{fig:timeline}
	\end{center}
\end{figure}

\section{Analytic model}\label{sec:methods}

In this work, we do not investigate how the objects observed at $z\approx6$ form. Instead, we start with the YMC (or YMC complexes) as observed and analyse the impact of their radiative feedback on the rest of the galaxy. We focus on the time before SN-driven winds from the young star cluster can destroy other dense gas clumps and quench star formation on a galaxy scale. 

Very young  (age $\lesssim 1-3\,\mathrm{Myr}$) star clusters can still be embedded in their birth cloud until a combination of radiation pressure, photoionization heating, and stellar winds stops further star formation and start to clear out the immediate surroundings \citep[see e.g.][and references therein]{krause_gas_2016}. When the radiation starts to leak out of the cluster birth cloud, radiative feedback reaches the peak of its impact on the rest of the host galaxy. As the cluster ages and the most massive stars explode as SNe, the total radiation field decreases, while the winds driven by SNe increasingly affect the galaxy (see Appendix~\ref{sec:timescales} for the relevant time-scales). Fig.~\ref{fig:timeline} illustrates the timeline of both radiative and SN feedback and highlights when each process is most important.

In the fiducial setup a young (age = 2 Myr\footnote{If the star cluster is embedded for longer (e.g. 4 Myr), the radiation field has decreased and we explore the dependence of our results on the age of the star cluster later (Fig.~\ref{fig:clump} in Sec.~\ref{sec:clumpy}).}) star cluster with $M_{\star} = 10^7\,\Msun$ irradiates pure hydrogen gas. In Sec.~\ref{sec:homogeneous} we start with the simple setup of a spherically symmetric, homogeneous gas cloud with the star cluster in its centre. By comparing the size of the Str\"omgren sphere to the expected extent of a typical $z=6$ galaxy, the maximum gas density below which the full galaxy can be ionized by the central star cluster can be derived. In Appendix~\ref{sec:timescales} we calculate the photoionization time-scale  and its dependence on the gas density to highlight the almost instantaneous formation of the Str\"omgren sphere compared to the age of the cluster.  

\begin{table}
	\centering
	\caption{Properties of the radiation fields from {\textsc{Starburst99}} from YMCs with a canonical IMF (cYMC) and a top-heavy IMF (tYMC). For reference the last column shows the very low metallicity (Z= 0.0004) radiation field using the Padova evolutionary tracks (see text). The important input parameters are listed in the top half: the initial star cluster mass ($M_{\star}$), the IMF slopes ($\alpha_1$, $\alpha_2$ and $\alpha_3$) for the indicated star mass intervals, and the metallicity $Z$. The resulting hydrogen ionizing luminosities $\dot{N}_{\ion{H}{I}}$ are listed in the bottom half for different cluster ages $t$.}\label{tab:overview}
	\begin{tabular}{lrrr} 
		\hline
			& cYMC & tYMC & low Z\\
		\hline
		\multicolumn{4}{l}{Input}\\
		\hline
		$M_{\star} \,[\Msun]$				& $10^7$ 		&$10^7$ 		&$10^7$\\
		$\alpha_1$  $(0.1 - 0.5\,\Msun)$ 	& 1.3			& 1.3			& 1.3 \\
		$\alpha_2$  $(0.5 - 1\,\Msun)$ 		& 2.3			& 2.3 		& 2.3\\
		$\alpha_3$  $(1 - 100\,\Msun)$ 		& 2.3			& 0.6 		& 0.6\\
		$Z$							& 0.002		& 0.002		& 0.0004\\
		\hline
		\multicolumn{4}{l}{Ionizing luminosities $\log \dot{N}_{\ion{H}{I}} \,[s^{-1}]$ at cluster age $t$}\\
		\hline
		$t$ = 2 Myr	& 53.7  & 54.8 & 53.8\\
		$t$ = 4 Myr	& 53.3  & 53.9 & 53.4\\
		$t$ = 6 Myr	& 52.7  & 53.0 & 52.9\\
		$t$ = 8 Myr	& 52.1  & 52.1 & 52.6\\
		$t$ = 10 Myr	& 51.5  & 51.4 & 52.2\\
		\hline
	\end{tabular}
\end{table}

In Sec.~\ref{sec:clumpy} we generalise the simple model to a clumpy ISM and calculate the maximum distance from the star cluster out to which a stable gas clump is fully ionized by the radiation of the YMC. We argue that radiation from one compact source, as claimed by e.g. \citet{bouwens_very_2017}, can prevent the formation of similarly compact sources within the same galaxy for low-mass haloes ($M_{200} \lesssim 10^{10}\,\Msun$\footnote{Halo masses throughout the paper are expressed as $M_{\mathrm{200}}$, which is defined as the mass of a spherical volume with radius $R_{\mathrm{200}}$ within which the mean density is 200 times the critical density at $z=6$.}). This process therefore increases the probability of observing only one dominant star cluster at any given time, which would naturally explain the inferred small intrinsic sizes of the galaxies of \citet{bouwens_very_2017} and the apparent change in the slope of the size-luminosity relation. 

The spectrum of the massive star cluster is calculated with the stellar evolution synthesis code {\textsc{Starburst99}} \citep{leitherer_starburst99:_1999, leitherer_effects_2014} for a simple stellar population forming instantaneously with a total initial stellar mass of $M_{\star} = 10^7\,\Msun$ and two different stellar initial mass functions (IMFs): a canonical IMF \citep[cIMF;][]{kroupa_variation_2001} with a high-mass slope of $\alpha=2.3$ and a very top-heavy IMF (tIMF) with a high-mass slope of $\alpha =0.6$. The latter yields the most extreme radiation field as it corresponds to birth densities assumed for ultra compact dwarf galaxies \citep{jerabkova_formation_2017} in the empirical model from \citet{marks_evidence_2012}.  A higher ionizing luminosity can also be the result of including binary stars in the spectral synthesis models. \citet{ma_binary_2016} and \citet{rosdahl_sphinx_2018} found a boost in the photon escape fraction in re-ionization simulations when including binary stars. The models cYMC (standard IMF, reference radiation field) and tYMC (top-heavy IMF, high radiation field) enclose the expected range of ionizing luminosities for a star cluster of $M_{\star} = 10^7\,\Msun$.

{\textsc{Starburst99}} models the integrated properties of stellar populations incorporating evolutionary tracks of individual stars. In version 7.0.1, the fiducial tracks from the Geneva group \citep{ekstrom_grids_2012, georgy_grids_2013} are available for two stellar metallicities \citep[sub-solar: Z = 0.002 and solar: Z = 0.014 with relative abundances from][]{asplund_chemical_2009} and two initial stellar rotational velocities: $v_{\mathrm{rot}} = 0$ and $v_{\mathrm{rot}}$ of 40 per cent of the break up velocity on the zero-age main sequence. 

We use the lower metallicity tracks as old globular clusters typically have sub-solar metallicities, e.g. -2.5<[Fe/H]<-0.3 in the GC sample of \cite{vandenberg_ages_2013}.\footnote{As we focus on the earliest stage of young star clusters, the exact choice of the stellar evolutionary track does not influence the main results. For a canonical IMF, the stellar evolution tracks from the Padova group for the lowest metallicity \citep[Z=0.0004,][]{fagotto_evolutionary_1994} result in an ionizing luminosity that is within 0.25 dex of the chosen Geneva track with Z = 0.002 for a cluster age of less than 6 Myr (Table~\ref{tab:overview}). Note that after 10 Myr, using the lowest metallicity Padova track returns a five times higher ionizing luminosity compared to our fiducial Geneva tracks. At that time, the ionizing luminosity used in our model can therefore be considered a lower limit for an extremely low metallicity YMC. }
Zero stellar rotation ($v_{\mathrm{rot}} = 0$) is assumed as the other model incorporates rotational velocities that might be too extreme \citep{leitherer_effects_2014} and does not reproduce the most massive stars very well \citep{martins_comparison_2013}.

{\textsc{Starburst99}} returns spectra and the \ion{H}{I} ionizing luminosities $\dot{N}_{\ion{H}{I}}\,[\mathrm{s}^{-1}]$ as a function of the age of the star cluster. $\dot{N}_{\ion{H}{I}}\,[\mathrm{s}^{-1}]$ is used for the analytic calculations while the spectra of the stellar populations serve as input for the spectral synthesis calculations with  {\textsc{Cloudy}} (see Sec.~\ref{sec:results}). Table~\ref{tab:overview} summarizes the input parameters for the YMC radiation fields cYMC and tYMC and lists $\dot{N}_{\ion{H}{I}}\,[\mathrm{s}^{-1}]$ for cluster ages $\leq 10\,\mathrm{Myr}$. The third column shows $\dot{N}_{\ion{H}{I}}\,[\mathrm{s}^{-1}]$ for an extremely low metallicity ($Z=0.0004$) stellar population to illustrate how the ionizing luminosity depends on the cluster metallicity.  

For gas whose recombination time-scale is long compared to the ionization time-scale (see Appendix~\ref{sec:timescales} for a discussion) the maximum gas mass that can be ionized until $3.3\,\mathrm{Myr}$ (time of first SNe) is $M_{\mathrm{max,i}} = m_{\mathrm{H}} \int_{\mathrm{t =0}}^{3.3\,\mathrm{Myr}} \dot{N}_{\ion{H}{I}}  \mathrm{d}t$ for pure hydrogen gas.  For the star cluster model cYMC (tYMC) $M_{\mathrm{max,i}} = 4.1 \times 10^{10}\,\Msun$ ($4.1 \times 10^{11}\,\Msun$). This gas mass is as high as the expected total gas mass in a $z=6$ galaxy hosted by an $M_{200}\gtrsim10^{12}\,\Msun$ dark matter halo according to the stellar mass - halo mass relation (and assuming a gas fraction of 50 per cent) from abundance matching \citep{behroozi_average_2013}. 

As the recombination time-scale decreases linearly with increasing density, the ionized gas mass falls below the theoretical $M_{\mathrm{max,i}}$ for high gas densities. In the following, we include the recombination rate and compare the ionized gas mass to the estimated mass of $z=6$ galaxies for both a homogeneous and a clumpy medium.

\subsection{Homogeneous medium}\label{sec:homogeneous}

Little is known about the gas distribution in faint, $z=6$ galaxies, but simulations suggest that they have a clumpy, irregular morphology \citep[see e.g.][]{ma_simulating_2018, trayford_star_2019}. In order to study the impact of radiative feedback on these galaxies, we decompose the galaxy into a volume-filling, homogeneous component and denser gas clumps embedded within the homogeneous medium. In this section we examine the maximum gas density for which the homogeneous inter-clump medium can be ionized throughout the whole galaxy. Afterwards, in Sec.~\ref{sec:clumpy}, the impact of the radiative feedback on the higher-density gas clumps is explored.

\subsubsection{Homogeneous gas sphere}\label{sec:homogeneoussphere}

First, we test whether the ionizing radiation of a young star cluster ($M_{\star} = 10^7\,\Msun$) is strong enough to ionize all gas within a galaxy in the most idealised case of a homogenous gas distribution before the SN delay time $\tau_{\mathrm{SN}}$ (Eq.~\ref{eq:SNtimescale}). The Str\"omgren sphere \citep[after][]{stromgren_physical_1939} describes the volume around an ionizing source where the ionization and recombination rates are in equilibrium. The Str\"omgren radius $R_{\mathrm {S}}$ determines the position of the boundary between ionized and neutral gas for a homogeneous density distribution, which we compare to the expected extent of the galaxy. $R_{\mathrm{S}}$ is given by

\begin{equation}\label{eq:RS}
	R_{\mathrm{S}} =  \left ( \frac{3}{4 \pi \alpha}  \frac{\dot{N}_{\ion{H}{I}}}{n_{\mathrm{e}}^2} \right )^{1/3}
\end{equation}

\noindent
and it depends on the rate of \ion{H}{I} ionizing photons $\dot{N}_{\ion{H}{I}}\,[\mathrm{s}^{-1}]$ emitted, the electron number density of the surrounding gas $n_{\mathrm{e}}\,[\mathrm{cm}^{-3}]$, and the recombination coefficient $\alpha \, [\mathrm{cm}^{3}\mathrm{s}^{-1}]$. In this toy model we assume the star cluster is embedded in a pure hydrogen gas cloud and that the gas within $R_{\mathrm{S}}$ is fully ionized. Therefore, the electron density is set to be equal to the hydrogen number density $\nH$. The case~B\footnote{We use $\alpha_B$ throughout the paper but as the setup is highly idealised the factor of 1.6 between $\alpha_A$ (case A recombination) and $ \alpha_B$ does not impact the interpretation of the results. }  (i.e. excluding recombination to the ground state) recombination coefficient for hydrogen at a temperature of $10^4\,\K$ is $\alpha = \alpha_B = 2.6 \times10^{-13}\,\mathrm{cm}^{3}\mathrm{s}^{-1}$ \citep{pequignot_total_1991}, which gives 

\begin{equation}\label{eq:RS2}
	R_{\mathrm{S}} = 1.5\,\,\mathrm{kpc}\, \left (  \frac{\dot{N}_{\ion{H}{I}}}{10^{53}\,\mathrm{s}^{-1}} \right )^{1/3} \left (  \frac{n_{\mathrm{H}}}{1\,\mathrm{cm}^{-3}} \right )^{-2/3} 
\end{equation}

\noindent
and a corresponding Str\"omgren mass of 

\begin{equation}\label{eq:MS2}
	M_{\mathrm{S}} = 3.2 \times 10^8 \,\mathrm{M}_{\odot}  \left (  \frac{\dot{N}_{\ion{H}{I}}}{10^{53}\,\mathrm{s}^{-1}} \right )   \left (  \frac{n_{\mathrm{H}}}{1\,\mathrm{cm}^{-3}} \right )^{-1}   .
\end{equation}

\begin{figure}
	\begin{center}
		\includegraphics[width=\linewidth, trim = 0cm 0cm 2.5cm 0cm]{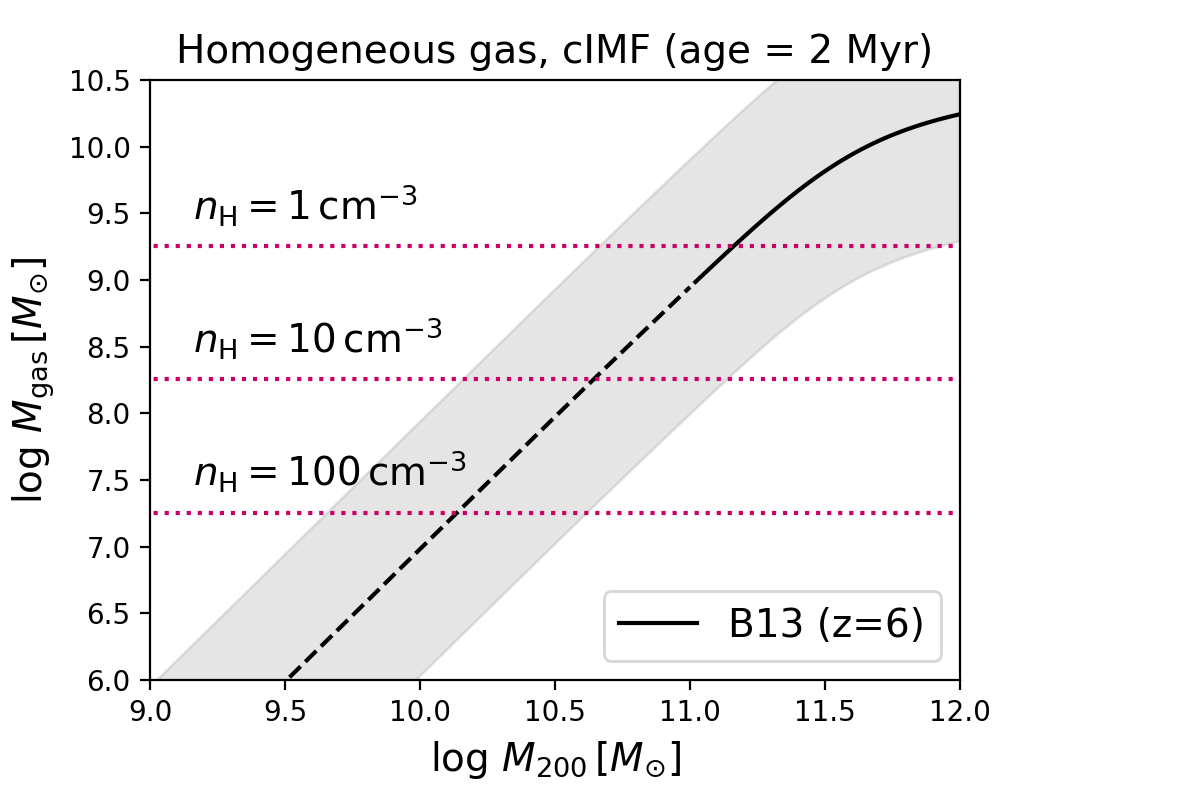}
		\caption{Str\"omgren mass (Eq.~\ref{eq:MS2}) for the radiation field of star cluster model cYMC at an age of 2 Myr ($\log \dot{N}_{\ion{H}{I}} [\mathrm{s}^{-1}] = 53.7$, Table~\ref{tab:overview}) and gas densities of $0.01$, $0.1$ and $1\,\ccm$ (horizontal dotted lines, as indicated). The black curve is the stellar mass (or gas mass for $f_g=0.5$) - halo mass relation from \citet{behroozi_average_2013} (B13) for $z=6$ (dashed line: extrapolation). The grey area indicates the gas mass - halo mass relation for galaxy gas fractions $f_g$ from 0.1 to 0.9. }
		\label{fig:stromgrenmass}
	\end{center}
\end{figure}

$\dot{N}_{\ion{H}{I}}$ varies with time (Table~\ref{tab:overview}) as the stellar population ages, but as the time-scales for both ionization and recombination are typically shorter ($< 1$ Myr, see Appendix~\ref{sec:timescales}) than the changes in $\dot{N}_{\ion{H}{I}}$, we use the equilibrium solution for a given $\dot{N}_{\ion{H}{I}}$. Fig.~\ref{fig:stromgrenmass} illustrates the Str\"omgren mass for an ionizing luminosity of $\log \dot{N}_{\ion{H}{I}} [\mathrm{s}^{-1}]= 53.7$ (model cYMC, at an age of 2 Myr, Table~\ref{tab:overview}) for different gas densities (as indicated). In order to compare the ionized gas mass to the total disc gas mass, the $z=6$ stellar mass - halo mass relation from \citet{behroozi_average_2013} is used. Depending on the gas fractions $f_g = M_{\mathrm{gas}} / (M_{\mathrm{gas}} + M_{\star})$ of the galaxy, the gas mass - halo mass relation lies within the grey region indicated in Fig.~\ref{fig:stromgrenmass} for $f_g$ between 0.1 and 0.9 with $f_g=0.5$ shown as a black curve. For a gas density of e.g. $\nH = 1\,\ccm$, the ionized gas mass exceeds the total gas mass of the galaxy for $\log M_{\mathrm{200}} [\Msun] \lesssim 11.2$ for $f_g = 0.5$.

\subsubsection{Homogeneous gas disc}\label{sec:homogeneousdisc}

So far, the gas density of the homogenous gas distribution is unconstrained, but with an estimate of the extent of the galaxy, the maximum gas density for which the YMC can ionize the full galaxy can be calculated. The compact objects in the sample of \citet{bouwens_very_2017} may be embedded in a larger, lower surface brightness disc, but the intrinsic size of the gaseous component of faint $z=6$ galaxies is largely unknown. \citet{huang_relations_2017} combined HST observations with abundance matching and derived a relation between the effective radius of galaxies ($R_{\mathrm{eff}}$) and the virial radius ($R_{\mathrm{200}}$) of their DM haloes for redshifts lower than 3. As their relation does not show a redshift dependence, we use their $R_{\mathrm{eff}}$ as a proxy for the size of the galaxy:

\begin{equation}\label{eq:Reff}
 	R_{\mathrm{eff}} = 1.68 \times \frac{\lambda}{\sqrt{2}} R_{\mathrm{200}} ,
 \end{equation}

\noindent
For a halo spin parameter $\lambda = 0.035$ \citep{tonini_measuring_2006} and a Hubble parameter of $H(z=6) = 700 \, \mathrm{km}\,\mathrm{s}^{-1}\,\mathrm{Mpc}^{-1}$ \citep{ade_planck_2016} the effective radius of a galaxy is approximated by 

\begin{equation}\label{eq:Reff2}
\begin{aligned}
	R_{\mathrm{eff}} = 0.4  \,\,\mathrm{kpc}\,  &  \left (  \frac{M_{\mathrm{200}}}{10^{10}\,\Msun}  \right )^{1/3}   \\
									& \left ( \frac{\lambda}{0.035} \right )  \left ( \frac{H(z)}{700 \, \mathrm{km}\,\mathrm{s}^{-1}\,\mathrm{Mpc}^{-1}} \right )^{-2/3} .
\end{aligned}
\end{equation}

\begin{figure}
	\begin{center}
		\includegraphics[width=\linewidth, trim = 0cm 0cm 2.5cm 0cm]{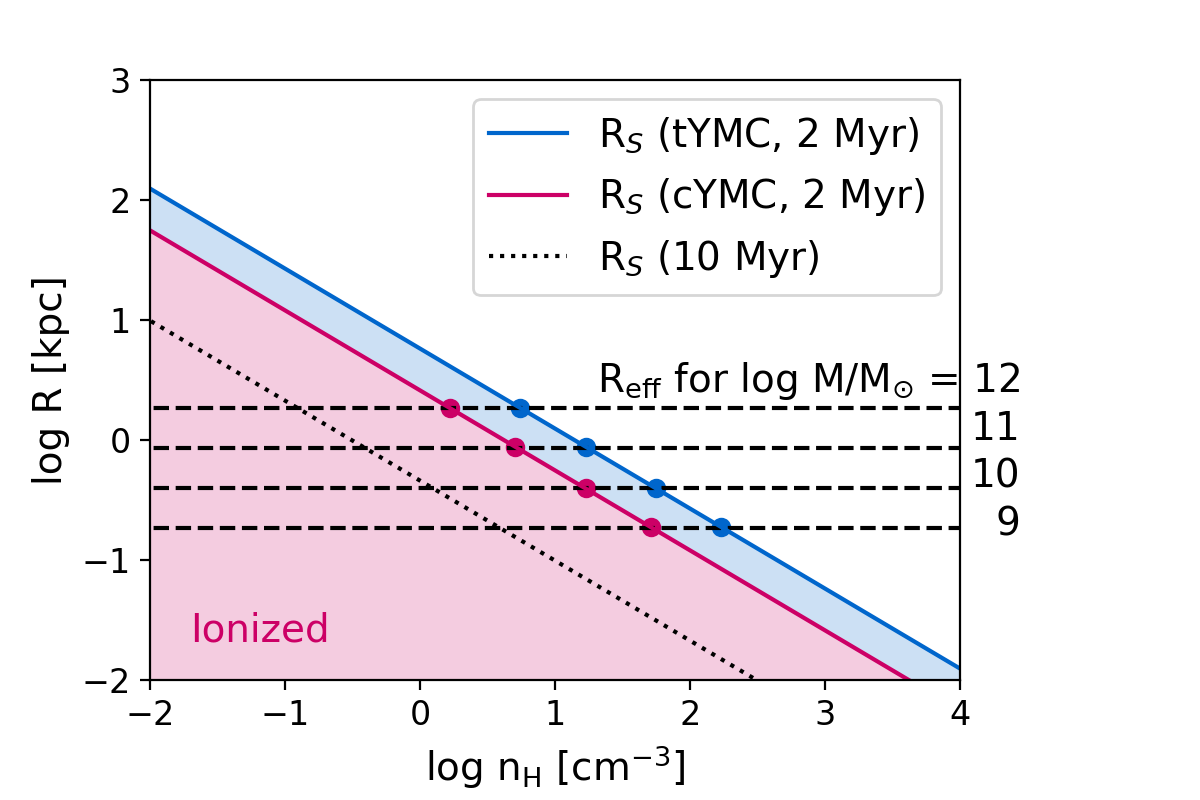}
		\caption{Str\"omgren radii (Eq.~\ref{eq:RS2}) for the ionizing luminosities of star clusters with an age of 2 Myr (red solid line: cYMC, blue solid line: tYMC). At a cluster age of 10 Myr, $\dot{N}_{\ion{H}{I}}$ is very similar for cYMC and tYMC (see Table~\ref{tab:overview}) and the Str\"omgren radius for both cases is indicated as black dotted line. The horizontal dashed lines are the estimated sizes $R_{\mathrm{eff}}$ (Eq.~\ref{eq:Reff2}) for galaxy discs at $z=6$ for halo masses of $\log M_{\mathrm{200}}[\Msun]$ between $9$ and $12$, as indicated. The points highlight the maximum density $n_{\mathrm{max}}$ (Eq.~\ref{eq:nmax}) for both radiation fields where the Str\"omgren radius exceeds the galaxy size.}
		\label{fig:stromgren}
	\end{center}
\end{figure}

A comparison between $R_{\mathrm{S}}$ and $R_{\mathrm{eff}}$ for the radiation fields of the star cluster models cYMC and tYMC at an age of 2 Myr is shown in Fig.~\ref{fig:stromgren}. The maximum density $n_{\mathrm{max}}$ for which all gas within $R_{\mathrm{eff}}$ is ionized by the YMC is indicated as points for each halo mass and follows from $R_{\mathrm{S}}(\nH,\dot{N}_{\ion{H}{I}}) = R_{\mathrm{eff}}(M_{\mathrm{200}})$:

\begin{equation} \label{eq:nmax}
	n_{\mathrm{H,max}} = 7 \,\, \ccm \,  \left ( \frac{\dot{N}_{\ion{H}{I}}}{10^{53}\,\mathrm{s}^{-1}}   \right  )^{1/2} \left (\frac{M_{\mathrm{200}}}{10^{10}\,\Msun}  \right)^{-1/2}  .
\end{equation}

\noindent
The maximum photoionized gas mass inside $R_{\mathrm{eff}}$ is $M_{\mathrm{gas,ion}} = 4 \pi / 3 R_{\mathrm{eff}}^3 \, n_{\mathrm{H,max}} \, m_{\mathrm H}$ or

\begin{equation}\label{eq:Mgasi}
	M_{\mathrm{gas,ion}} = 5\times10^7\,\Msun \left ( \frac{\dot{N}_{\ion{H}{I}}}{10^{53}\,\mathrm{s}^{-1}}   \right  )^{1/2} \left (\frac{M_{\mathrm{200}}}{10^{10}\,\Msun}  \right)^{1/2} .
\end{equation}

\noindent
Comparison of $M_{\mathrm{gas,ion}}$ to the total gas mass expected in the disc allows us to estimate the halo mass range for which the full galaxy is ionized in this toy model. 

If the galaxy disc were thin, only a small fraction of the emitted photons would reach the gas in the disc and therefore the ionized gas mass $M_{\mathrm{S,disc}}$ would be smaller than $M_S$ from the assumption of spherical symmetry. 
The ionized gas mass in this case is reduced by two spherical caps, where we assume that any gas that exists in these caps above or below the disc is so tenuous that its contribution to the total gas mass is negligible. 

As an estimate for the vertical extent of the gas disc we use the Jeans length $\lambda_J$. This assumes a self-gravitating disc that is stabilised by an isotropic pressure (e.g. thermal or turbulent) that can be expressed by an effective temperature $T$. The Jeans length is defined as

\begin{equation}\label{eq:jeanslength}
	\lambda_J = \left ( \frac{\gamma k_B T}{G \mu m_H^2 \nH} \right )^{1/2}
\end{equation}

\noindent
or

\begin{equation}\label{eq:jeanslength2}
	\lambda_{\mathrm{J}}  = 1.14 \,\, \mathrm{kpc} \, \left ( \frac{T}{10^4\,\K} \right )^{1/2}  \left ( \frac{\nH}{1 \,\ccm} \right )^{-1/2}
\end{equation}

\noindent
where $\gamma = 5/3$ is the ratio of specific heats, $k_B$ is the Boltzmann constant, $\mu=1$ the mean particle mass for neutral hydrogen, $m_{\mathrm H}$ is the mass of the hydrogen atom, $\nH$ is the hydrogen number density of the gas and $T=10^4\K$ is a typical value for the temperature of the warm neutral medium.

\begin{figure}
	\begin{center}
		\includegraphics[width=\linewidth]{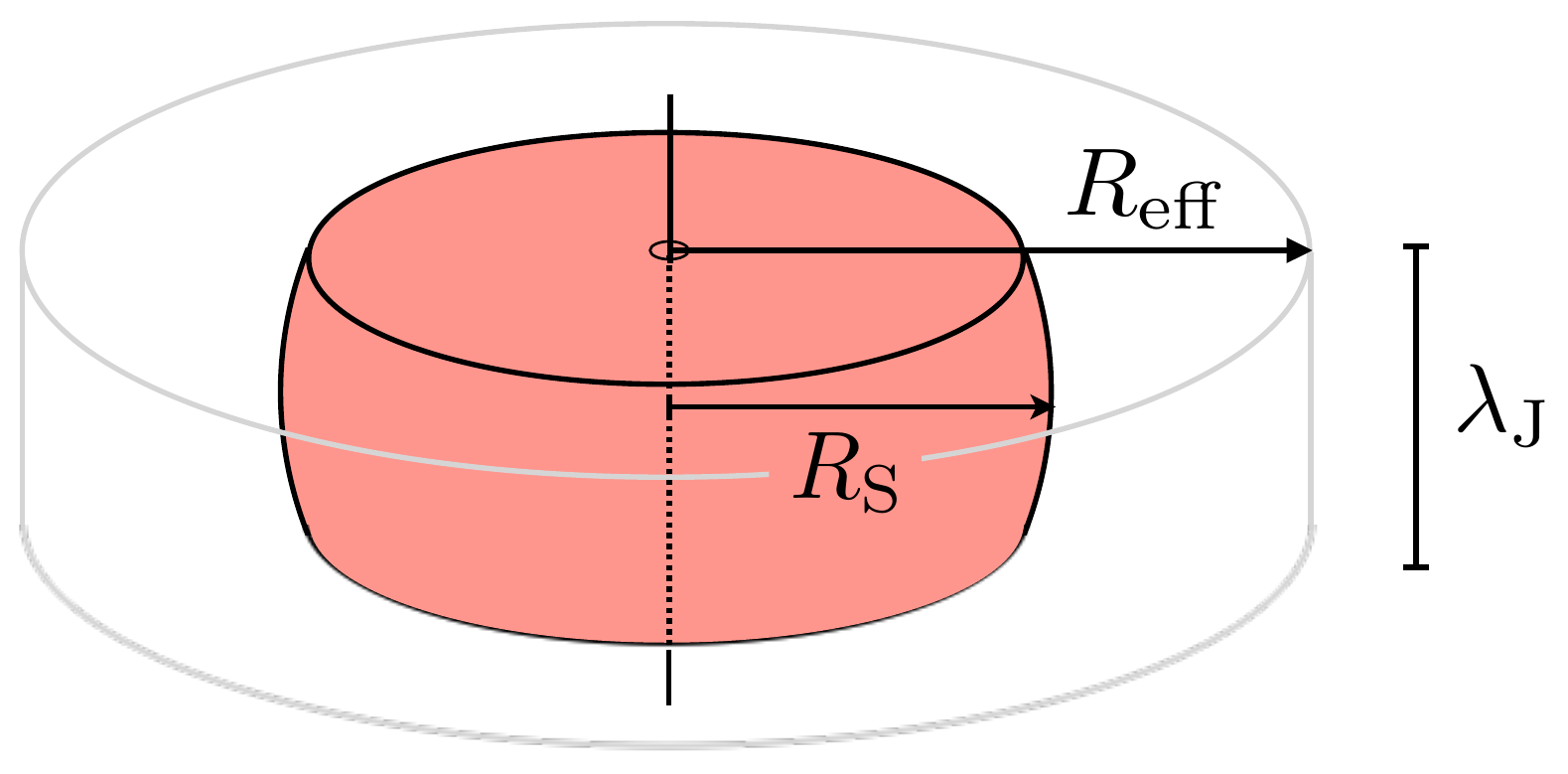}
		\caption{Schematic view of the geometry of the ionized gas mass distribution (with radius $R_{\mathrm{S}}$; shaded volume) within a disc of radius $R_{\mathrm{eff}}$ and thickness $\lambda_{\mathrm{J}}$ (grey outlines) as assumed in Sec.~\ref{sec:homogeneousdisc}. Note that while the maximum density $n_{\mathrm{max}}$ (Eq.~\ref{eq:nmax}) for which gas can be ionized until $R_{\mathrm{eff}}$ is the same as for a spherically symmetric gas distribution, the gas mass $M_{\mathrm{S, disc}}$ is reduced depending on the thickness of the disc (Eq.~\ref{eq:MSdisc1}). }
		\label{fig:schematiccaps}
	\end{center}
\end{figure}

Fig.~\ref{fig:schematiccaps} shows a schematic plot of the ionized gas mass in a homogeneous gas disc with thickness $\lambda_{\mathrm{J}}$ and radius $R_S$. In this case the ionized gas mass in the disc ($M_{\mathrm{S, disc}}$) is reduced compared to the spherical $M_S$, depending on the thickness of the gas disc:

\begin{equation}\label{eq:MSdisc1}
	\frac{M_{\mathrm{S,disc}}}{M_{\mathrm{S}} } = \frac{1}{2} \left  ( \frac{\lambda_{\mathrm{J}}/2}{R_{\mathrm{S}}} \right )  \left [  3 -  \left ( \frac{\lambda_{\mathrm{J}}/2}{R_{\mathrm{S}}}\right )^2 \right ]  .
\end{equation}

\noindent
For  $R_{\mathrm{S}} = R_{\mathrm{eff}}$ (and therefore $n = n_{\mathrm{max}}$), the ratio $\lambda_{\mathrm{J}}/2$ to $R_{\mathrm{S}}$ is

\begin{figure}
	\begin{center}
		\includegraphics[width=\linewidth, trim = 0cm 0cm 2.5cm 0cm]{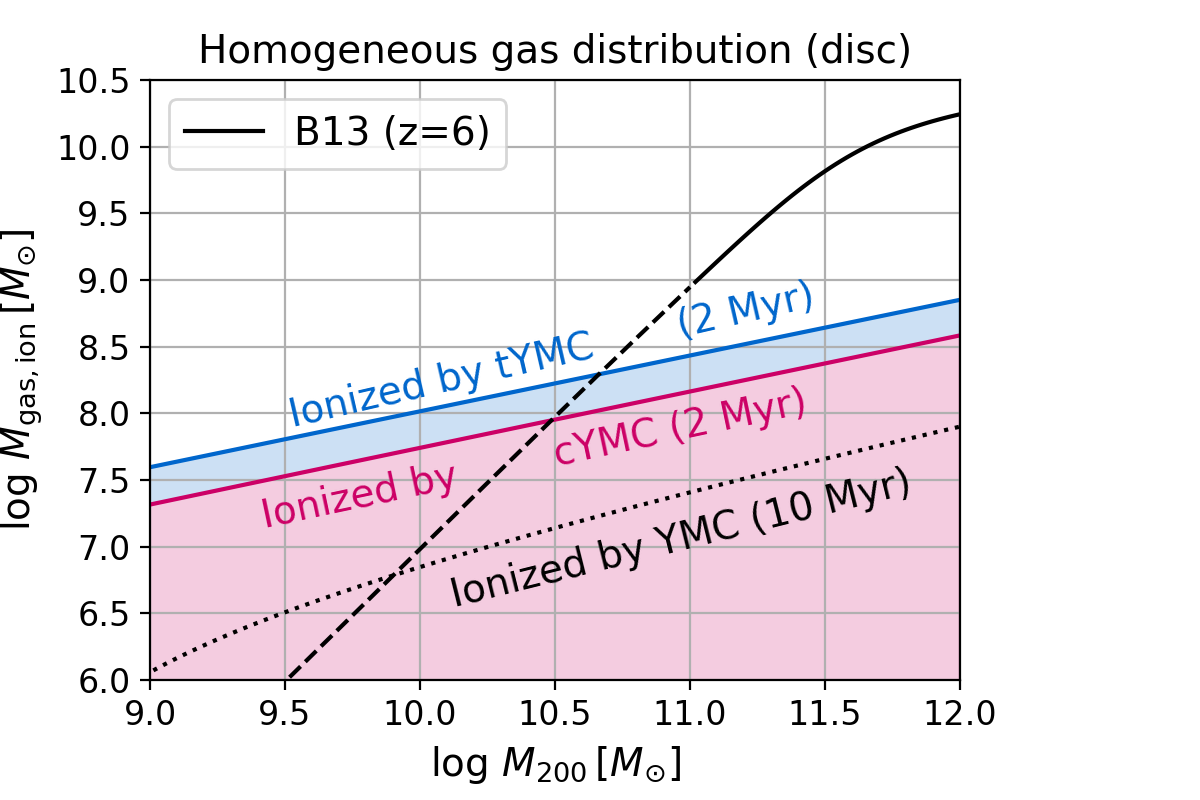}
		\caption{Maximum ionized gas mass ($M_{\mathrm{gas,ion}}$, Eqs.~\ref{eq:MSdisc1} and \ref{eq:MSdisc2}) within $R_{\mathrm{eff}}$ for a disc of thickness $\lambda_{\mathrm{J}}$ (see schematic in Fig.~\ref{fig:schematiccaps}). The black curve is the stellar mass - halo mass relation from \citet{behroozi_average_2013} (B13) for $z=6$ (dashed line: extrapolation). Assuming a gas fraction of 50 per cent, the full galaxy is ionized by the respective radiation field in this toy model for halo masses $M_{\mathrm{200}}$ where $M_{\mathrm{gas,ion}}$ exceeds the B13 relation}
		\label{fig:fbarcap}
	\end{center}
\end{figure}

\begin{equation}\label{eq:MSdisc2}
	\frac{\lambda_{\mathrm{J}}/2}{R_{\mathrm{S}}} = 0.54 \left ( \frac{T}{10^4\,\K} \right )^{1/2} \left ( \frac{\dot{N}_{\mathrm{H}}}{10^{53}\,\mathrm{s}^{-1}} \right )^{-1/4} \left ( \frac{M_{\mathrm{200}}}{10^{10}\,\Msun} \right )^{-1/12}
\end{equation}

\noindent
and the resulting maximum ionized gas mass $M_{\mathrm{S, disc}}$ for the cluster models cYMC and tYMC (at a cluster age of 2 and 10 Myr, as labelled) is shown in Fig.~\ref{fig:fbarcap}. For the fiducial values, the assumed disc is very thick, with a diameter ($2R_{\mathrm{S}}$) less than a factor of two larger than its thickness $\lambda_{\mathrm{J}}$ (Eq.~\ref{eq:MSdisc2}).

For halo masses $\log M_{\mathrm{200}} [\Msun] \lesssim 10.5$ ($\lesssim 10.7$) the gas mass that can be ionized by cYMC (tYMC) is larger than the baryonic mass from the stellar-mass halo-mass relation for $z=6$ by \citet{behroozi_average_2013}. This indicates that the radiation field of a YMC can be strong enough to ionize a full low-mass galaxy. In higher-mass haloes ($M_{\mathrm{200}} \gtrsim 10^{10.7}\,\Msun$), the maximum ionized gas mass does not exceed the estimated gas mass within $R_{\mathrm{eff}}$ and one YMC could not globally suppress star formation within the galaxy. The general message is therefore: we expect a halo mass, below which the ionizing luminosities of the assumed YMCs can ionize more gas than the total gas mass in homogeneous, volume-filling low-density gas expected in these galaxies.

\subsection{Clumpy medium}\label{sec:clumpy}

We have established in Sec.~\ref{sec:homogeneous} that homogeneous, low-density gas can be ionized by the YMC throughout the whole galaxy. In a more realistic high-redshift galaxy, the gas is expected to exhibit an irregular and clumpy morphology. We therefore explore in this section how the radiation field ionizes gas clumps with different masses and distances to the YMC. The maximum distance between the YMC and a gas clump is estimated by $R_{\mathrm{eff}}$ (Eq.~\ref{eq:Reff2}), the assumed total extent of the gas distribution.

\begin{figure}
	\begin{center}
		\includegraphics[width=\linewidth, trim = 2.cm 7.cm 8.8cm 0.7cm, clip]{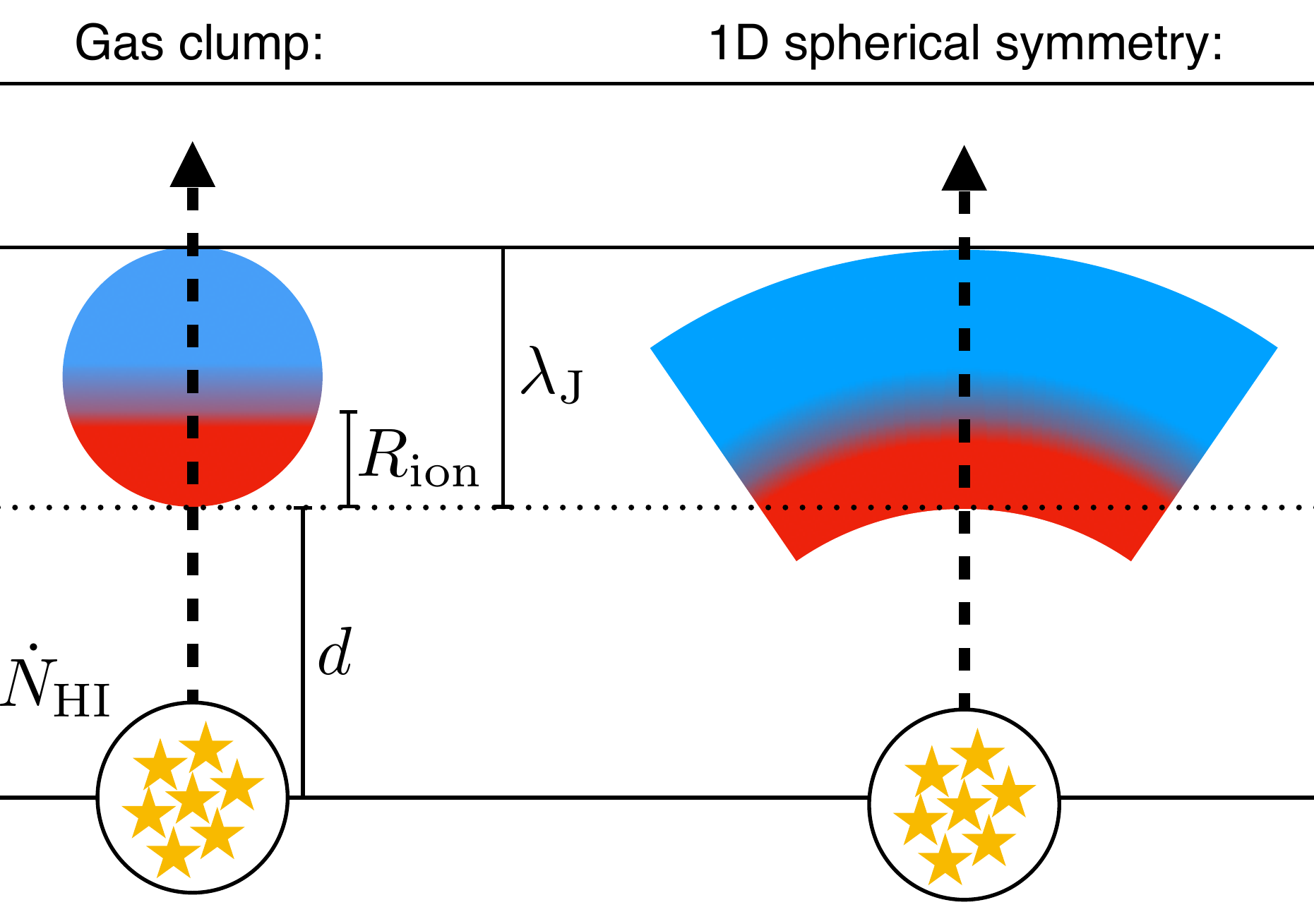}
		\caption{Schematic of the clumpy medium setup discussed in Sec.~\ref{sec:clumpy}. A gas clump with an extent close to its Jeans length $\lambda_{\mathrm{J}}$ gets irradiated by a massive star cluster with an ionizing luminosity $\dot{N}_{\ion{H}{I}}$ at a distance $d$. The gas clump gets photoionized on the illuminated side of the gas cloud until a depth of $R_{\mathrm{ion}}$ where the electron fraction $n_e / n_H = 0.5$ (left panel). We define the clump to be fully ionized if $R_{\mathrm{ion}} > \lambda_{\mathrm{J}}$. The right panel illustrates the assumption of spherically symmetric gas shells.}
		\label{fig:sketch}
	\end{center}
\end{figure}

In a clumpy galaxy, the radiation from the star cluster needs to heat or even ionize the embedded dense gas regions to prevent star formation in the galaxy. Whether a clump can self-shield from the radiation field and remain cold depends on the gas density as well as the distance between the clump and the ionizing source. The sketch on the left hand side of Fig.~\ref{fig:sketch} shows a gas clump at a distance $d$ to a star cluster emitting ionizing photons at a rate of $\dot{N}_{\ion{H}{I}}$.  If the gas clumps are embedded in low-density gas, we assume that the largest absorption occurs in the densest gas clump and neglect the absorption through the lower-density gas. This is especially justified in the case where the surrounding medium is ionized and therefore optically thin to the hydrogen ionizing radiation. We discuss the contribution of a volume filling lower-density gas in Appendix~\ref{sec:absorption} but find its contribution negligible for our setup. 

We assume the Jeans length $\lambda_J$ (Eq.~\ref{eq:jeanslength2}) as a typical clump size as it is a characteristic length scale for a self-gravitating gas \citep[e.g.][]{schaye_model-independent_2001}.
As illustrated in Fig.~\ref{fig:sketch} (left), the illuminated side of the gas clump is ionized first while the opposite side of the cloud stays neutral if it is shielded from the stellar radiation. If the depth of the ionization front $R_{\mathrm{ion}}$ within the gas cloud is larger than the cloud size (here: $\lambda_{\mathrm{J}}$ with $T = 10^4\,\K$ as the most conservative choice for initially neutral gas), then the gas clump with density $n$ is fully ionized by the radiation field from the source with luminosity $\dot{N}_{\ion{H}{I}}$ at distance $d$.

We focus on the 1D line-of-sight through the centre of the cloud (left panel of Fig.~\ref{fig:sketch}) to analytically formulate the position of $R_{\mathrm{ion}}$ inside the gas clump. For the following comparison between the extent of the gas cloud and the position of the ionization front, the 1D line-of-sight through the centre of the gas clump is assumed to be spherically symmetric about the position of the ionizing source (right side of Fig.~\ref{fig:sketch}). This allows us to describe the depth of the ionization front $R_{\mathrm{ion}}$ by a Str\"omgren sphere. While this assumption allows us to formulate the analytic toy model, the resulting $R_{\mathrm{ion}}$ for individual 1D line-of-sights through a given column density of gas (here: $n \lambda_{\mathrm{J}}$) are independent of the underlying shape of the gas cloud.

For a clump or shell with constant density, $R_{\mathrm{ion}}$ can be determined analytically by calculating for which $R_{\mathrm{ion}}$ the volume of the shell $V_{\mathrm{shell}} = 4 \pi / 3 \left [  (d + R_{\mathrm{ion}})^3  - d^3 \right ]$ equals the volume of the Str\"omgren sphere $V_{\mathrm{S}} = 4 \pi / 3 R_{\mathrm{S}}^3$ for the gas density of the clump. Here $d$ is the distance between the inner edge of the clump or shell and the ionizing source (see Fig.~\ref{fig:sketch}).
Solving for $R_{\mathrm{ion}}$ in $V_{\mathrm{shell}}  = V_{\mathrm{S}}$ results in:

\begin{equation}\label{eq:Ri}
	R_{\mathrm{ion}} = R_{\mathrm{S}} \cdot f(c)  
\end{equation}

\noindent
where

\begin{equation}\label{eq:Ri2}
	 f(c)  =  \left ( \sqrt[3]{1+c^3} - c \right ) 
\end{equation}

\noindent
with $c \equiv d / R_{\mathrm{S}}$. In the next step we use $R_{\mathrm{S}}$ from Eq.~\ref{eq:RS2} and $R_{\mathrm{ion}} = \lambda_{\mathrm{J}}$ (with $\lambda_{\mathrm{J}}$ from Eq.~\ref{eq:jeanslength2}) to formulate the maximum distance $d_{\mathrm{max}}$ to the ionizing source, where a gas clump with density $n$ (again assuming that $n \approx n_e \approx \nH$) is completely ionized:

\begin{equation}\label{eq:dclump}
d_{\mathrm{max}} \, [\mathrm{kpc}]  = \sqrt{A \left ( \frac{\dot{N}_{\ion{H}{I}}}{10^{53}\,\mathrm{s}^{-1}} \right )  n^{-1.5} - Bn^{-1}} - Cn^{-0.5}
\end{equation}

\noindent
with $A = 0.91$, $B = 0.108$, and $C = 0.57$. This defines the distance up to which gas clumps are ionized as a function of their gas density $n$ and the ionizing luminosity $\dot{N}_{\ion{H}{I}}$.   

\begin{figure*}
	\begin{center}
		\includegraphics[width=\linewidth, trim = 3.5cm 0cm 2cm 0cm, clip]{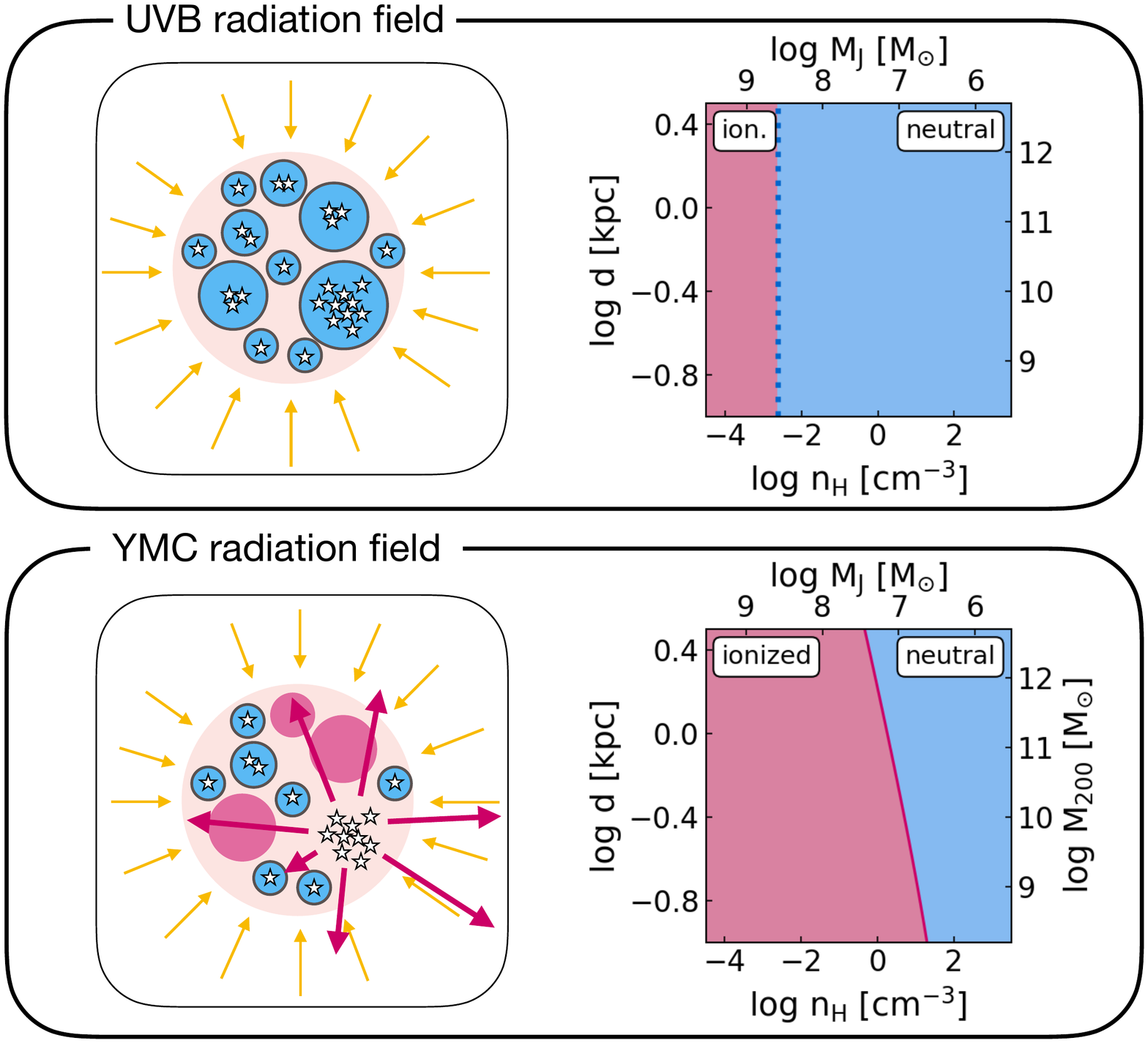}
		\caption{ Left panels: Sketch of a clumpy galaxy as discussed in Sec.~\ref{sec:clumpy} without (top) and with (bottom) a YMC. Blue circles symbolise neutral gas clumps while red circles represent ionized gas clumps. Right panels: A clump with density $\nH$ (bottom axis) at distance $d$ from the YMC (left y-axis) is either fully ionized ($d < d_{\mathrm{max}}$, Eq.~\ref{eq:dclump} or $n < n_{\mathrm{max}}$ for the UVB case; red region) or remains partly neutral ($d \ge d_{\mathrm{max}}$; blue region) for the UVB (top panel) or the fiducial cYMC radiation field (age = 2 Myr, $M_{\star} = 10^7\,\Msun$, bottom panel). The right y-axis in the right panels gives the halo mass $M_{\mathrm{200}}$ for which the maximum distance (left y-axis) equals the estimated size of the galaxy (Eq.~\ref{eq:Reff2}). With only the UVB radiation field, self-gravitating gas clumps with densities $\log \nH [\ccm]> -2.6$ can self-shield and potentially cool to form star clusters. These densities correspond to Jeans masses of up to $\approx 4 \times 10^8\,\Msun$ (top axis, top right panel). As we neglect shielding by the inter-clump medium, the UVB solution is independent of the position within the galaxy (see Appendices~\ref{sec:absorption} and \ref{sec:HM12} for details). After the YMC has formed and cleared out its birth cloud, the strong ionizing radiation can also keep gas clumps with higher densities warm and ionized. In haloes with masses below the value indicated by the right y-axis, the radiation can ionize such clumps throughout the galaxy. Only smaller clumps with $M_{\mathrm{J}} \lesssim 10^7\,\Msun$ can self-shield due to their higher densities (bottom right panel).  }
		\label{fig:sketchradfield}
	\end{center}
\end{figure*}

The bottom right panel in Fig.~\ref{fig:sketchradfield} shows $d_{\mathrm{max}}$ (left y-axis) for various gas densities $n$ (bottom x-axis). For comparison, the maximum gas density for which a clump with size $\lambda_{\mathrm{J}}$ can be fully ionized by the UV background (UVB) radiation field at $z = 6$ \citep[][hereafter:~HM12]{haardt_radiative_2012} is shown in the top right panel in Fig.~\ref{fig:sketchradfield}. For the HM12 radiation field, a plane-parallel geometry is assumed (see Appendix~\ref{sec:HM12} for details) and the limiting density is therefore independent of the distance.

The maximum distances $d_{\mathrm{max}}$ are related to the halo mass $M_{\mathrm{200}}$ (right y-axis) via the estimated size of galaxies $d = R_{\mathrm{eff}}$ (Eqs.~\ref{eq:Reff}, \ref{eq:Reff2}). If therefore a clump with density $n$ can be ionized at a distance $d \ge R_{\mathrm{eff}}$, then we assume that gas clumps of this density will be ionized by the radiation field throughout the whole galaxy. 
Furthermore, the left y-axis is scaled to the Jeans mass $M_{\mathrm{J}} = 4\pi/3( \lambda_{\mathrm{J}}/2)^3 n\, m_{\mathrm {H}} = 1.9\times 10^7\,\Msun n^{-1/2}$ for a gas temperature of $10^4\,\K$ and the respective gas density $n [\ccm]$ (top x-axis). 

While dense ($n \gtrsim 100 \,\ccm$) self-gravitating gas clumps with low Jeans masses ($M_{\mathrm{J}}\approx 10^6\,\Msun$) can remain neutral within the galaxy for all halo masses considered, massive gas clumps ($M_{\mathrm{J}} > 10^7\,\Msun$) remain ionized thanks to their lower gas densities. 
For example, a YMC with a mass of $10^7\,\Msun$ and an age of 2 Myr can keep a gas clump with a Jeans mass of $M_{\mathrm{J}} \ge 10^{7}\,\Msun$ ionized throughout the whole galaxy in haloes with $\log M_{\mathrm{200}} [\Msun] \lesssim 10.5$. In higher-mass haloes, these clumps can remain neutral, as the galaxy sizes exceed the limiting distance $d_{\mathrm{max}}$.

Summarising, self-gravitating massive gas clumps ($\log M_{\mathrm{J}} [\Msun] \approx 8.5$ for $T=10^4\,\K$) can cool to below 100~K\footnote{see Fig.~\ref{fig:clumpstoptempHM12} for the numerical results of the thermal state of gas clumps exposed to HM12.} if the only heating source is the  \citetalias{haardt_radiative_2012} UVB radiation field for $z=6$ (Fig.~\ref{fig:sketchradfield}, top right panel). If these massive structures collapse and form a $10^7\,\Msun$ star cluster, the radiation field from the YMC can keep adjacent gas clumps with low densities and therefore large Jeans masses ionized and warm ($\approx 10^4\,\K$) up to distances $d_{\mathrm{max}}$ that can exceed the estimated sizes of low-mass (expressed in terms of $M_{\mathrm{200}}$) galaxies. The sketch in Fig.~\ref{fig:sketchradfield} illustrates this idea. 
This process has to work fast, as the YMC radiation field at a cluster age of 10 Myr is already comparable to the \citetalias{haardt_radiative_2012} radiation field\footnote{This is a conservative estimate as the ionizing luminosity of star clusters with lower metallicities decreases more slowly with cluster age and is therefore still higher after 10 Myr (a factor of 5 for $Z = 0.0004$ compared to $Z=0.1\Zsol = 0.00134$, see Table~\ref{tab:overview}).}. 

In galaxies with small disc sizes and therefore limited maximum distances to the YMC, gas clumps with large Jeans masses can be ionized throughout the whole disc. Only higher-density and hence, given the assumption that the clump mass equals the Jeans mass, lower-mass gas clumps self-shield and remain (partially) neutral. In large discs, as is typical for more massive haloes (Eq.~\ref{eq:Reff2}), gas clumps in the disc can self-shield if they are at sufficiently large distances to the YMC. Similar to the results in Sec.~\ref{sec:homogeneous}, a maximum halo mass (and therefore maximum disc size) is expected below which the YMC can ionize all gas clumps in a galaxy. 

\paragraph*{Star cluster mass, IMF and age dependence:}
The lines in Fig.~\ref{fig:clump} show $d_{\mathrm{max}}$ for different IMFs (top panels: cYMC, bottom panels: tYMC) as well as different YMC masses (left panels) and YMC ages (right panels). If the YMC needs more than 2 Myr to clear out its birth cloud, the weaker radiation field at an age of 4 Myr (dashed line, right panels in Fig.~\ref{fig:clump}) can ionize Jeans masses of $M_{\mathrm{J}} \ge 10^{7}\,\Msun$ in slightly less massive haloes of $\log M_{\mathrm{200}} [\Msun] \lesssim 9$ (cYMC) or $\lesssim 11$ (tYMC). After 8 Myr (dash-dotted line), clumps with Jeans masses of up to $\approx 10^8\,\Msun$ remain neutral in all haloes (both tYMC and cYMC), which is still $\approx 4$ times smaller compared to the case where the only radiation field is the \citetalias{haardt_radiative_2012} background at $z=6$ (Fig.~\ref{fig:sketchradfield}, top right panel). While the stronger radiation field of more massive YMCs can ionize also smaller gas clumps, we highlight that the assumed IMF plays a critical role. The ionizing luminosity of cYMC with a mass of $10^7\,\Msun$ is comparable to that of tYMC with a mass of $10^6\,\Msun$ (both at an age of 2 Myr, see also Table~\ref{tab:overview}).

\section{Numerical parameter exploration}\label{sec:results}

\begin{figure}
	\begin{center}
		\includegraphics[width=\linewidth, trim = 0cm 0.5cm 0.5cm 0.4cm, clip]{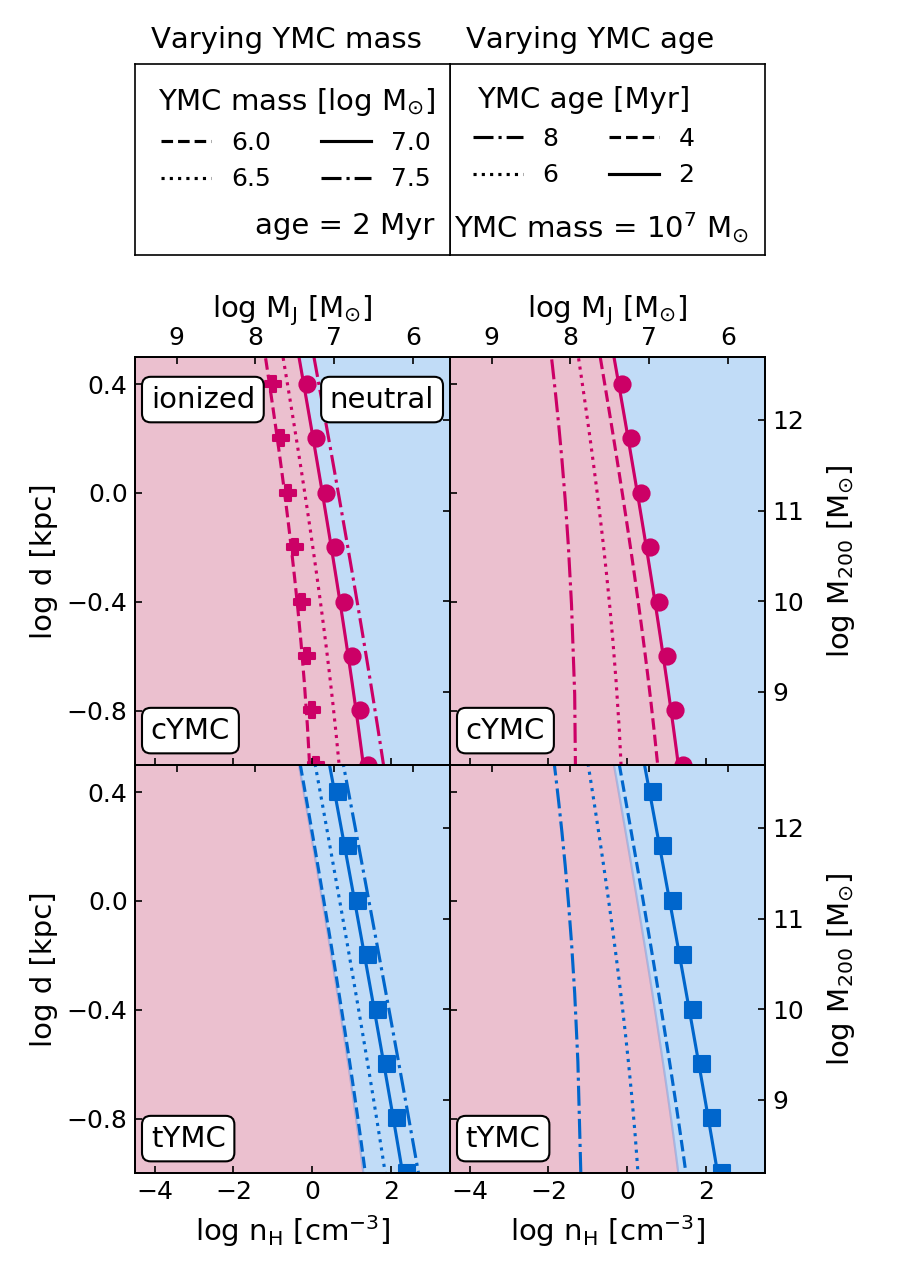}
		\caption{As in the bottom right panel of Fig.~\ref{fig:sketchradfield} but for varying YMC masses at a constant YMC age (2 Myr, left panels) and varying YMC ages at a constant YMC mass ($10^{7}\,\Msun$, right panels) for two different IMFs: cYMC (top row) and tYMC (bottom row). For some reference models, the results from \textsc{Cloudy} are over-plotted with datapoints (see Sec.~\ref{sec:results} for details). The coloured regions  correspond to the parameter space where gas clouds are either ionized ($d < d_{\mathrm{max}}$; red region) or remain neutral ($d \ge d_{\mathrm{max}}$; blue region) for the fiducial model (cYMC, 2 Myr, $10^{7}\,\Msun$), repeated from the bottom right panel of Fig.~\ref{fig:sketchradfield}. In the same way, each line separates the parameter space into ionized (towards lower densities and smaller distances) and neutral (towards higher densities and larger distances) for each radiation field.}
		\label{fig:clump}
	\end{center}
\end{figure}

In this section we compare the results from Sec.~\ref{sec:methods} to 1D radiative transfer simulations. This allows us to explore the dependence of the results on the gas metallicity and dust content, as the analytic models in Sec.~\ref{sec:methods} assume pure hydrogen gas and only distinguish between fully ionized and fully neutral gas. For more detailed information about the thermal state of the neutral gas beyond the ionization front and the effect of metals and dust, we use numerical results from {\textsc{Cloudy}} \citep[][here: version 17.00]{ferland_cloudy_1998, ferland_2013_2013, ferland_2017_2017}. {\textsc{Cloudy}} is a spectral synthesis and radiative transfer code that makes use of extensive atomic and molecular databases to calculate the physical and chemical conditions of gas in astrophysical environments. After specifying a spectral shape as well as the luminosity or intensity of a radiating source (we use the spectra from models cYMC and tYMC), {\textsc{Cloudy}} calculates absorption and emission features for an adaptive number of zones through a slab or shell of gas. As in the analytic calculations, the constant density case is used, while now the gas temperature, ion abundances and electron densities are iterated to their thermal equilibrium state for each zone. 

For comparison with the analytic Str\"omgren radius from Eq.~\ref{eq:RS2}, we define the numerical Str\"omgren radius, $R_{\mathrm{0.5}}$, as the depth into the gas, where the ratio between the electron number density $n_{\mathrm{e}}$ and the hydrogen number density $\nH$ is $n_{\mathrm{e}}/\nH = 0.5$. 

For a clumpy gas distribution, we set up a 2D grid of {\textsc{Cloudy}} simulations where the varied parameters are the distance, $d$ (see the left panel in Fig.~\ref{fig:sketch}), between the ionizing source and the gas clump with density $n$. The {\textsc{Cloudy}} calculations return the ion fractions of hydrogen as well as the gas temperature as a function of depth into the gas clump. We discuss the temperature profile in Sec.~\ref{sec:preheating} and focus here on the depth into the cloud, where $n_{\mathrm{e}}/\nH = 0.5$, defined as $R_{\mathrm{0.5}}$ (parallel to the definition of $R_{\mathrm{ion}}$ in Fig.~\ref{fig:sketch}). Together with the characteristic length scale of a self-gravitating cloud, the Jeans scale $\lambda_\mathrm{J}$, the depth $R_{\mathrm{0.5}}$ is used as a measure of how much of the gas clump is (mostly) ionized ($n_{\mathrm{e}}/\nH > 0.5$). 

As before, a constant temperature of $10^4\,\K$ is assumed for the Jeans length $\lambda_\mathrm{J}$. Beyond the ionization front at $R_{\mathrm{0.5}}$ the temperature decreases, reducing the Jeans length of the gas clump ($\lambda_\mathrm{J} \propto T^{1/2}$). Since a smaller gas clump of a given density is easier to ionize, the constant $10^4\,\K$ temperature assumption is the most conservative choice to investigate the effect of ionizing radiation onto these gas clumps.

Throughout this section, the values for solar metallicity ($\Zsol = 0.0134$) and the elemental abundance ratios are taken from \citet{asplund_chemical_2009} (for $Z = \Zsol$: $X = 0.738$, $Y = 0.249$). For {\textsc{Cloudy}} runs with dust, we use the ``orion'' grain set which includes both graphitic and silicate grains (see the {\textsc{Cloudy}}  documentation for details). Grains are destroyed mainly in the shock waves of SN blast waves \citep[e.g.][]{jones_grain_1996}. As we explicitly focus on the time before the SN shock passes through the galaxy, and the temperatures of $\approx 10^4\,\K$ are too low for thermal sputtering to be efficient \citep{tielens_physics_1994}, we assume a constant dust-to-metal ratio of 0.42 also within the ionized bubble and do not include any grain destruction processes in these {\textsc{Cloudy}} simulations.

As a first test, the analytic results from Fig.~\ref{fig:sketchradfield} and Fig.~\ref{fig:clump} are compared to the numerical results from {\textsc{Cloudy}} for low metallicity ($Z=0.1\,\Zsol$), dust-free gas. This allows us to test the assumptions from the analytic model (i.e. pure hydrogen gas, constant temperature, $T=10^4\,\K$, and constant electron number density, $n_e = \nH$). 

The limiting distance (or density) between ionized and (partly) neutral is shown in Fig.~\ref{fig:clump} for the analytic solution from Eq.~\ref{eq:dclump} (solid lines) and the {\textsc{Cloudy}} runs (symbols) for ionizing luminosities from star clusters of different IMFs, ages and masses. Both the shape and the normalization of the data points match the theoretical predictions very well.   

\subsection{Gas metallicity and dust dependence}\label{sec:metallicity}

Faint galaxies at $z=6$ are expected to have very low metallicities and, assuming a constant dust-to-metal ratio, also very little dust. The analytic calculations as well as the fiducial numerical model represent this scenario. In more massive galaxies or in galaxies at lower redshifts, the higher dust content provides an additional shielding channel against radiation from nearby star clusters. Here, we explore how metals and dust impact the results of the dust and metal-free analytic model. This allows us to shed some light on the systematic differences between radiative feedback in dust-free (i.e. faint, high-$z$) vs. dusty (i.e. more massive, low-$z$) galaxies in the proposed scenario.

\paragraph*{Homogeneous medium:} \label{sec:numhomogeneous}

\begin{figure}
	\begin{center}
		\includegraphics[width=\linewidth, trim = 2.2cm 0.7cm 3cm 1cm, clip]{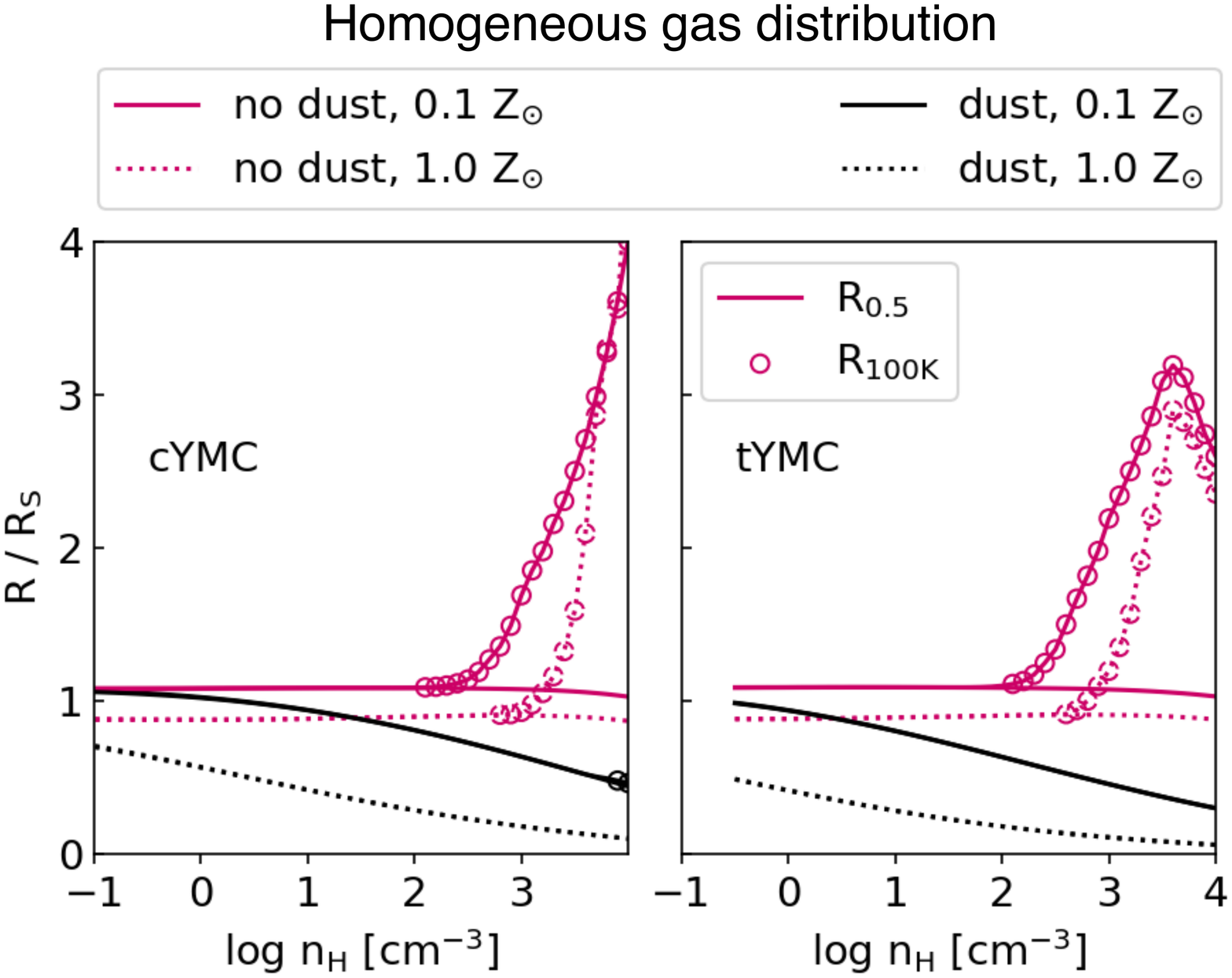}
		\caption{The ratio between $R_{\mathrm{0.5}}$, the numerically determined radius where $n_{\mathrm{e}} / \nH = 0.5$, and $R_{\mathrm{S}}$, the analytic radius of the Str\"omgren sphere (Eq.~\ref{eq:RS2}), is displayed with solid (gas metallicity $0.1\,\Zsol$) and dotted (gas metallicity $1\,\Zsol$) lines. Lines with symbols show the ratio between the radius where the gas temperature drops below 100~K ($R_{\mathrm{100K}}$) and $R_{\mathrm{S}}$.  $R_{\mathrm{100K}}/R_{\mathrm{S}}$ is only plotted for densities where $R_{\mathrm{100K}} \ne R_{\mathrm{0.5}}$.  Red lines indicate the results for dust-free gas, while for the black lines a constant dust-to-metal ratio of 0.42 is assumed.
		Results are shown for model cYMC (left panel) and tYMC (right panel), both for a cluster age of 2 Myr. }
		\label{fig:stromgrencomp}
	\end{center}
\end{figure}

The results for a radiating source surrounded by a medium with constant density are summarised in Fig.~\ref{fig:stromgrencomp} for model cYMC (left panel) and tYMC (right panel), both for the radiation field at a cluster age of 2 Myr. Here, we focus on the lines without symbols ($R_{\mathrm{0.5}}$) while the meaning of the lines with symbols ($R_{\mathrm{100K}}$) will be explained in Sec.~\ref{sec:preheating}.

For gas without dust and at a low metallicity of $0.1\,\Zsol$, $R_{\mathrm{0.5}}$ is within a few percent of the analytic solution, $R_{\mathrm{S}}$ (red solid line). Increasing the metallicity to $\Zsol$ decreases $R_{\mathrm{0.5}}$ systematically by around 10 per cent (red dotted line). Therefore, the gas metallicity alone does not have a large effect on the size of the ionized bubble. 
Including dust grains (black lines) leads to smaller $R_{\mathrm{0.5}}$ at high densities (see Fig.~\ref{fig:stromgrencomp}): for gas densities of $\log \nH [\ccm] = 4$,  $R_{\mathrm{0.5}}$ is reduced by a factor of 2 (10) compared to the dust-free model for a metallicity of $0.1\,\Zsol$ ($\Zsol$).
The volume that can be ionized by a given radiation field is therefore considerably reduced in dusty gas with solar metallicity compared to the gas with very little dust content.

\paragraph*{Clumpy medium:}

\begin{figure}
	\begin{center}
		\includegraphics[width=\linewidth, trim = 0cm 0.5cm 0cm 2cm, clip]{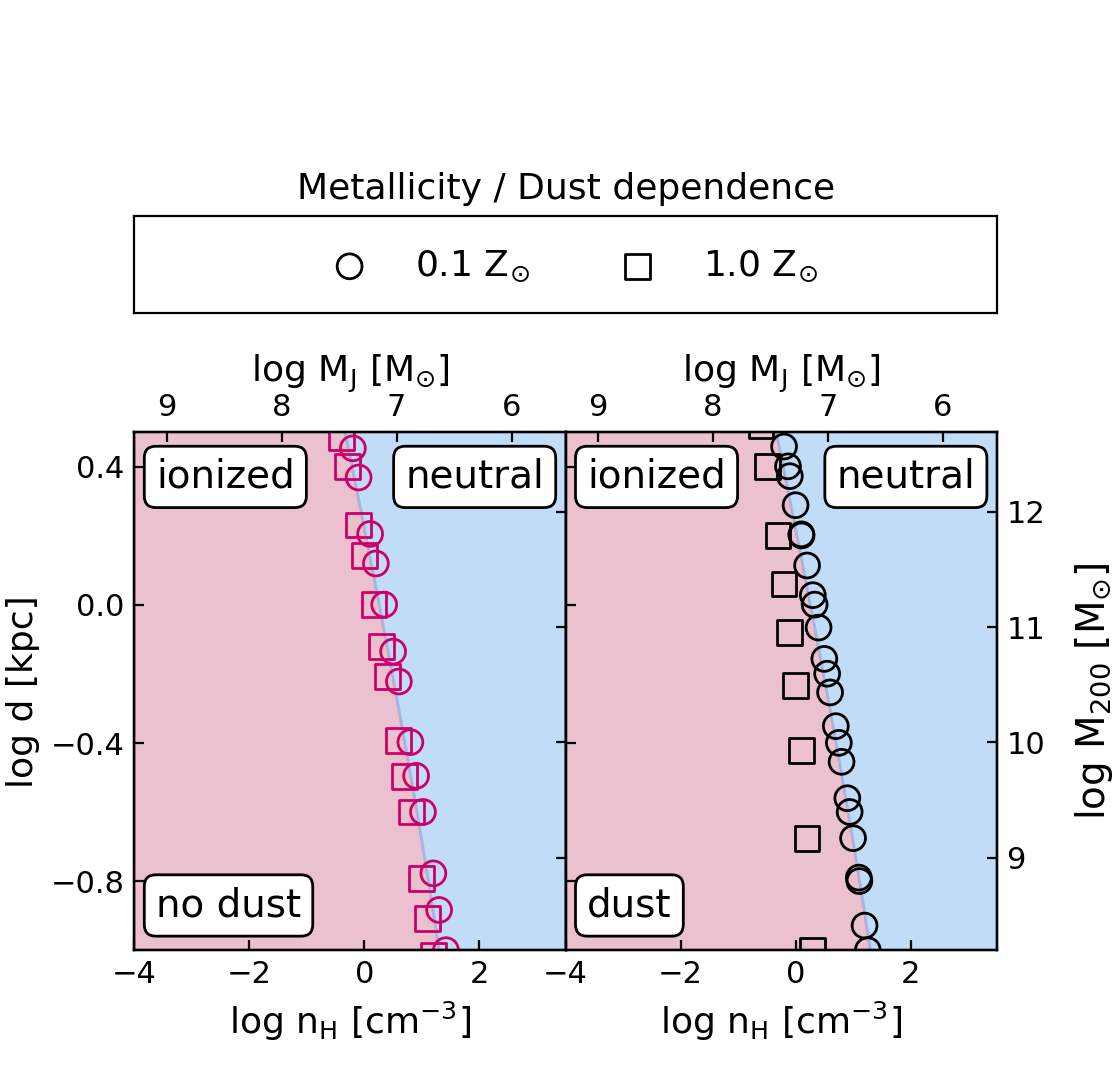}
		\caption{As Fig.~\ref{fig:clump} but for two different gas metallicities (circles: $0.1\,\Zsol$, squares: $\Zsol$) and including dust with a constant dust-to-metal ratio of 0.42 (right panel, black symbols). All results are for the fiducial radiation field (cYMC, $t=2\,\mathrm{Myr}$, $10^{7}\,\Msun$) and the coloured regions from the analytic result are repeated from Fig.~\ref{fig:clump} for reference.}
		\label{fig:clumpsolar}
	\end{center}
\end{figure}

We illustrate in Fig.~\ref{fig:clumpsolar} the dependence of the results from Fig.~\ref{fig:sketchradfield} on dust content and metallicity. Both low metallicity cases (with and without dust), as well as the dust-free solar metallicity case scatter closely around the analytic solution for pure hydrogen gas (solid line). On the other hand, dust clearly enhances the self-shielding of the gas for $Z = \Zsol$ and moves the limit between (fully) ionized clump and (partly) neutral clumps to somewhat lower gas densities. For dusty gas with $Z = \Zsol$, a gas clump with a Jeans mass of $M_{\mathrm{J}} = 10^7\,\Msun$ remains neutral in this radiation field, even in very low-mass haloes ($\log M_{\mathrm{200}} [\Msun] < 9$).

For galaxies with a low dust content, as expected for faint $z=6$ galaxies, radiative feedback from the same YMC can ionize larger fractions of its host galaxy. This suggests that the galaxy-wide suppression of star formation from radiative feedback plays a larger role at high redshifts than in the local Universe. 

As the numerical results for dust-free gas for metallicities of $0.1\Zsol$  and $\Zsol$ are very close to the analytic, metal-free solution, the results are expected to be valid also for metallicities below $0.1\Zsol$.

\subsection{Pre-heating}\label{sec:preheating}

Photons with energies for which the ionization cross section of atomic hydrogen is zero (e.g. optical light, or UV below 13.6 eV) can pass deeper into the gas than hydrogen ionizing radiation and can heat (mostly neutral) gas beyond $R_{\mathrm{0.5}}$. 
This could be an additional mechanism for suppressing star formation without ionizing the entire gas cloud. 
We use $R_{\mathrm{100K}}$ as a measure for the importance of pre-heating of neutral gas, which is defined as the depth into the gas, where the temperature drops below $100\,\K$. We chose this low temperature rather than e.g. $1000 \K$ to illustrate the upper limit of this effect.

\paragraph*{Homogeneous medium:}  
For an estimate of how much neutral gas can be heated to temperatures of $>100\,\K$, the ratio $R_{\mathrm{100K}}/R_{\mathrm{S}}$ is plotted in Fig.~\ref{fig:stromgrencomp} (marked with symbols where $R_{\mathrm{100K}} \ne R_{\mathrm{0.5}}$). For most densities, the lines for $R_{\mathrm{100K}}$ and $R_{\mathrm{0.5}}$ overlap, which means that the temperature drops steeply to below 100~K at the ionization front. 

\begin{figure}
	\begin{center}
		\includegraphics[width=\linewidth, trim = 0.2cm 1cm 0.5cm 1.0cm, clip]{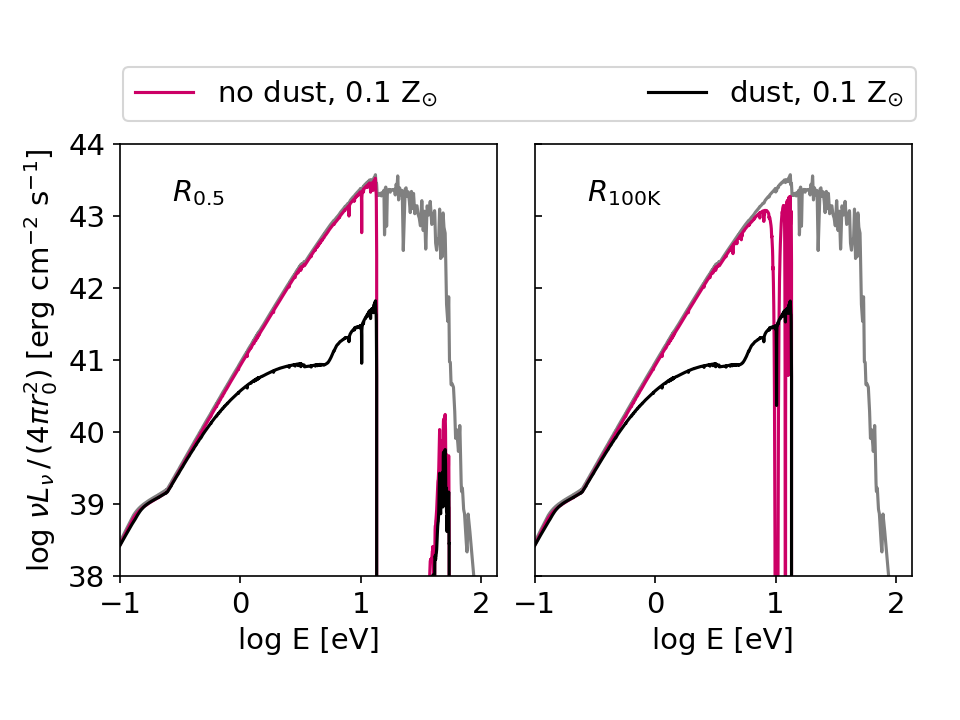}
		\caption{Incident spectrum (grey line in both panels) for cYMC at a cluster age of 2 Myr. Transmitted spectra are shown for gas densities of $\log \nH [\ccm] = 3.5$ at $R_{\mathrm{0.5}}$ (left panel) and $R_{\mathrm{100K}}$ (right panel) for low metallicity ($Z=0.1\,\Zsol$) gas without dust (red lines) and with dust grains (black lines). At the ionization front ($R_{\mathrm{0.5}}$, left panel), radiation with energies between $\sim 1$ eV and the hydrogen ionization energy of $13.6 \,\mathrm{eV}$ is heavily attenuated by dust. All spectra are normalized to $4\pi r_0^2$ where $r_0  = 10^{18}\mathrm{cm}$ is the inner radius of the cloud. }
		\label{fig:spectrumcomp}
	\end{center}
\end{figure}

Pre-heating is only significant for dust-free gas with high gas densities ($\nH \gtrsim 500\ccm$), where the radius at which the temperature drops below 100~K can be a few times larger than $R_{\mathrm{S}}$. For gas that contains dust grains (dust-to-gas mass ratio of $5.6\times10^{-3}$ for $\Zsol$), the radiation field does not heat up the gas beyond the ionization front. Fig.~\ref{fig:spectrumcomp} illustrates the difference in the spectra of model cYMC for an individual density ($\log \nH [\ccm]= 3.5$) at $R_{\mathrm{0.5}}$ (left panel) and $R_{\mathrm{100K}}$ (right panel). For dusty gas, even at a low metallicity of $0.1\,\Zsol$, the radiation field at energies between $\approx 1$ and $13.6$ eV, is drastically attenuated at $R_{\mathrm{0.5}}$. The dust-free gas in comparison is barely absorbing at these energies at $R_{\mathrm{0.5}}$. Deeper into the cloud, at $R_{\mathrm{100K}}$ (right panel), the Lyman series is visible as absorption features in the dust-free gas, but the optical and UV radiation field is still up to 2 dex stronger compared to the dusty gas case\footnote{The used Starburst99 spectra do not contain X-ray radiation.}. 

Therefore, in addition to the larger volume of the ionized region in dust-free gas, as shown in Sec.~\ref{sec:metallicity}, the neutral gas beyond the ionization front is also warmer in low-metallicity, dust-free gas, compared to gas with a dust content typical for the solar neighborhood. The main pre-heating mechanism here is the photoionization of excited states of hydrogenic species (label $\mathrm{Hn=2}$ in {\textsc{Cloudy}}). As illustrated in Fig.~\ref{fig:spectrumcomp} photons with energies that can excite hydrogen atoms (i.e. $10.2 \le E [\mathrm{eV}] < 13.6$) are very efficiently absorbed by dust grains. This leads to a lower fraction of excited hydrogen atoms and subsequently to a lower heating rate compared to dust-free gas.

\paragraph*{Clumpy medium:}

Even without fully ionizing a gas clump, the strong radiation field could still increase the thermal equilibrium temperature for a large fraction of its volume. For a homogeneous density distribution we showed in Fig.~\ref{fig:stromgrencomp} (symbols) that the pre-heating radius, $R_{\mathrm{100K}}$, can be a factor of a few larger than $R_{\mathrm{0.5}}$, which traces the location of the ionization front. For individual gas clumps $R_{\mathrm{100K}}$ and $R_{\mathrm{0.5}}$ are in the following compared to the estimated size of the gas clump to assess the fraction of the clump that is affected by pre-heating. 

\begin{figure}
	\begin{center}
		\includegraphics[width=\linewidth, trim = 2.2cm 2.2cm 2.5cm 2.5cm, clip]{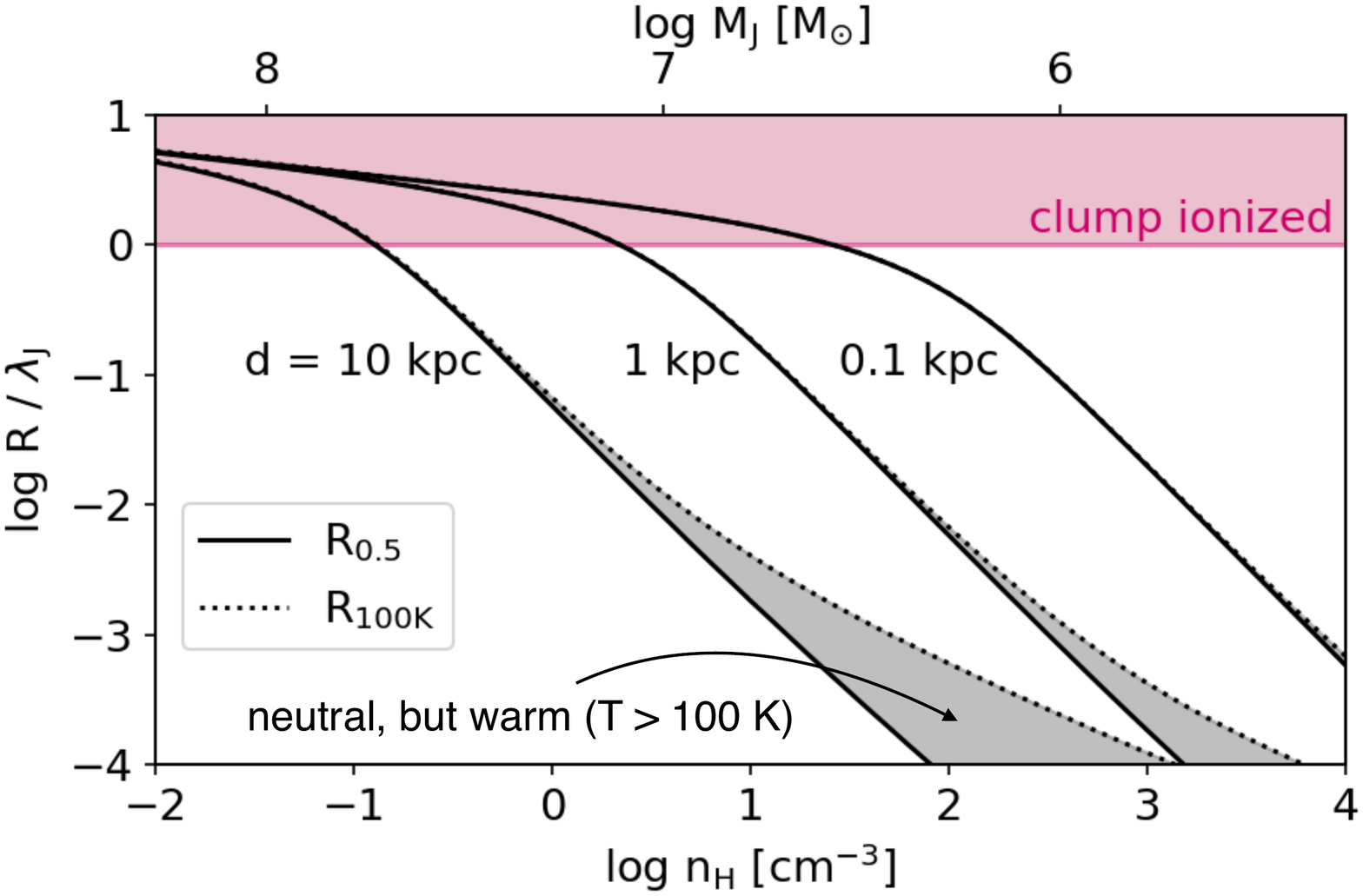}
		\caption{Depth into the low-metallicity ($0.1 \Zsol$), dust-free gas clump within which $n_{\mathrm{e}}/\nH>0.5$ ($R_{\mathrm{0.5}}$, solid lines) and within which the gas temperature is higher than $100\,\K$ ($R_{\mathrm{100K}}$, dotted lines). The depths are normalised to the assumed size of the clump ($\lambda_{\mathrm{J}}$ for $T = 10^4\,\K$) and the inner edge of the gas clump is located at distances of $d = 10, 1, 0.1\,\mathrm{kpc}$ (indicated). The gas clump is completely ionized where $R_{\mathrm{0.5}} / \lambda_{\mathrm{J}} \ge 1$ (red region). For $R_{\mathrm{100K}}>R_{\mathrm{0.5}}$ mostly neutral gas beyond the ionization front is pre-heated to $T>100\K$ by the radiation field (grey areas). The top x-axis shows the Jeans mass $M_{\mathrm{J}}$ of a self-gravitating gas clump with density $\nH$ and a temperature of $10^4\K$ for reference.}
		\label{fig:clumpstoptemp}
	\end{center}
\end{figure}

For a cYMC (age = 2 Myr) in low-metallicity ($0.1 \Zsol$) and dust-free gas, Fig.~\ref{fig:clumpstoptemp} shows $R_{\mathrm{0.5}}$ (solid lines) and $R_{\mathrm{100K}}$ (dotted lines), both relative to the assumed size of the clump ($\lambda_{\mathrm{J}}$) for distances of $d = 0.1,\,1,\,10\,\mathrm{kpc}$. Gas is mostly neutral, but heated to $> 100\,\K$ where $R_{\mathrm{100K}}>R_{\mathrm{0.5}}$ (indicated as grey areas).

Heating of gas beyond the ionized bubble only occurs for densities where $R_{\mathrm{100K}} / \lambda_{J} \ll 1$ (Fig.~\ref{fig:clumpstoptemp}). The depth into the clump where the gas is heated to a temperature of $>100\,\K$ is therefore much smaller than the assumed size of the clump. This means that the situation where the radiation field heats up the full gas clump to temperatures above 100~K without ionizing it, does not occur in this setup. Typically, the self-shielded part of the cloud is cold ($T<100\,\K$). Only for high densities (see Fig.~\ref{fig:clumpstoptemp}) can a thin shell between the ionized outskirts and the cold core be both neutral and warm ($T>100\,\K$).

We conclude that the radiation feedback is most effective, both in terms of ionizing and heating gas, when the metallicity is low and, even more importantly, the dust content is low\footnote{The reduced opacity of dust-poor gas, as is expected for low-mass high-$z$ galaxies, also leads to a decrease in radiation pressure on the gas compared to the Galactic metallicity and dust content. While this effect is not explored here, this could be an explanation for the higher total masses of dust- and metal-poor YMCs \citep{howard_universal_2018}.}. In that case, the radiation from YMCs can heat high-density ($\log \nH [\ccm] \gtrsim 3$) gas to temperatures above 100~K at distances that are a factor of a few larger than the ionization radius. 

\section{Discussion}\label{sec:discussion}

We showed that radiative feedback from an individual YMC can indeed be powerful enough to fully ionize homogeneous gas in galaxy discs of low-mass haloes, assuming that the radiation escapes the birth cloud on a short time-scale (a few Myr). In this strong radiation field, gas clumps within the ionized gas can self-shield only for densities that correspond to Jeans masses that are too low to form similarly massive YMCs, especially for dust-free gas. It is therefore expected that a large fraction of ionizing photons can leave the thick galaxy disc when this process is effective. This would predict a high escape fraction $f_{\mathrm{esc}}$ of Lyman continuum (LyC) photons for these faint compact objects at $z=6$.

Direct observations of LyC photons produced during or shortly after re-ionization to measure $f_{\mathrm{esc}}$ is impossible due to their small mean free paths  at these redshifts (e.g. $\lambda_{\mathrm{mfp}}^{13.6\mathrm{eV}} \le 20\,\mathrm{Mpc}$ for $z>4.5$; \citetalias{haardt_radiative_2012}; see \citealp{dayal_early_2018} for a recent review on early galaxy formation). Promising candidates for Lyman leaking systems at low redshift ($z\approx0.3$) are ``Green Peas'' \citep{cardamone_galaxy_2009}, a class of compact starburst galaxies first identified in GalaxyZoo \citep{lintott_galaxy_2008}. Especially the Green Peas with the highest ($>5$) $O_{\mathrm{32}} =$ [\ion{O}{III}]$\lambda 5007$/[\ion{O}{II}]$\lambda3727$ ratios have measured escape fractions from 8 per cent \citep{izotov_eight_2016} up to 72 per cent \citep{izotov_low-redshift_2018}. For example, J1154+2443 is  a compact star-forming disc galaxy at $z = 0.37$ with a Lyman continuum escape fraction of 46 per cent \citep{izotov_j1154+2443:_2018}: It has a half-light radius of $\approx 0.2 \,\mathrm{kpc}$, a low metallicity of $\approx 0.1\,\Zsol$ ($12+\log O/H = 7.65$), a total (old and young stellar population) stellar mass of $\log M_{\star} [\Msun] = 8.2$, a star formation rate of $18.9\,\Msun \, \mathrm{yr}^{-1}$, and a starburst age of $2-3\,\mathrm{Myr}$. These measured escape fractions are very high compared to an average of $f_{\mathrm{esc}} \le 2$ per cent measured in low-redshift galaxies with lower $O_{\mathrm{32}}$ ratios (e.g. \citealp{borthakur_local_2014, Leitherer_direct_2016}; see also \citealp{puschnig_lyman_2017}). 

For higher redshifts ($z\approx 2-4$), low escape fractions of a few per cent have been derived from $\gamma$-ray bursts \citep{chen_new_2007}, deep UV imaging \citep[e.g.][]{grazian_lyman_2016, steidel_keck_2018}, and in a sample of Ly$\alpha$ and H$\alpha$ emitters in the COSMOS field \citep{matthee_production_2017}.
Despite the low average values for $f_{\mathrm{esc}}$, individual objects can have very high escape fractions (e.g. ``Ion2'' \citealp{vanzella_peering_2015} with a LyC escape fraction of $64^{+110}_{-10}$ per cent \citealp{de_barros_extreme_2016}, see also \citealp{vanzella_hubble_2016}), similar to the low-redshift examples.

Based on the individual measurements at $z\approx0.3$ and $z \approx 3$, the highest values for $f_{\mathrm{esc}}$ seem to occur preferentially in starburst galaxies with a very young (age $\approx$ 2-3 Myr), compact stellar component, low metallicities, and low stellar masses ($\log M_{\star} [\Msun] \approx 8-9$ with higher $f_{\mathrm{esc}}$ for lower $M_{\star}$; \citealp{izotov_low-redshift_2018}). 

Therefore, if the faint compact objects observed at $z=6$ are as metal- and dust-poor as expected from high-$z$ mass-metallicity relations \citep[e.g. fig 5. in][]{mannucci_lsd:_2009}, their low stellar masses and high UV luminosities make them good candidates for high-$z$ analogues of the Lyman leaking galaxies observed at lower redshifts. In this case, the observed compact sources are bright star clusters that outshine their host galaxy. As observations only recently started detecting this new class of objects, their dust content and gas properties are currently unknown. Future observations with the James Webb Space Telescope can provide more detailed information, including measurements of the galaxy colours, metallicities, and dust content. With this information, it will be possible to refine and test the models presented here.

\section{Summary}\label{sec:summary}

In this work, we showed that radiation feedback from a compact, massive, star-forming region at $z=6$ can prevent the formation of additional young massive clusters (YMCs) in low-mass galaxies. This may explain the lack of additional bright clusters in these objects. Supernova feedback has a delay time that is much longer than the formation time of a star cluster, and only becomes important after the host star cluster has already faded, but radiation feedback affects the rest of the galaxy almost instantly after the YMC has cleared out its birth cloud. This early radiative feedback can qualitatively explain why the measured sizes of faint galaxies at redshift 6 are much smaller than expected from the extrapolated luminosity-size relation of brighter galaxies. It also offers an explanation for the question raised in \citet{bouwens_very_2017} on the likelihood of having only one dominant star cluster in a faint galaxy.    

The model is based on a gas-rich, low-mass galaxy in which a perturbation (e.g. tidal shock, mergers, gas inflow, instabilities) leads to the formation and compression of massive gas clumps.
As long as the only heating source is the UV background (UVB), a self-gravitating gas cloud with an initial temperature of $\approx 10^4\,\K$ can self-shield and subsequently cool to $\ll10^4\,\K$ for densities $\log \nH [\ccm] <  2.6$ (Fig. ~\ref{fig:clumpstoptempHM12}). This density corresponds to an initial (for $T=10^4\,\K$) Jeans mass of $\log M_{\mathrm{J}} [\Msun] \approx 8.5$ (top right panel of Fig.~\ref{fig:sketchradfield}). If the collapse of this massive gas cloud leads to the formation of a YMC, the radiation field within the galaxy increases drastically as soon as the YMC has cleared out its birth cloud. Because of the higher ionization rate, gas with higher densities can be ionized and hence, gas clumps with Jeans masses that could self-shield and cool in the UVB case, now remain ionized and warm (bottom right panel of Fig.~\ref{fig:sketchradfield}). 
For example, for the fiducial model\footnote{Radiation field from a star cluster with a canonical IMF (cYMC), an age of 2 Myr and a mass of $M_{\star}  =10^7\,\Msun$}, a gas clump with a Jeans mass of $M_{\mathrm{J}} = 10^7\,\Msun$ is fully ionized out to distances that correspond to the disc sizes expected for galaxies within haloes of $\log M_{\mathrm{200}}[\Msun] \approx 10.3$ (bottom right panel of Fig.~\ref{fig:sketchradfield}).

At ages of $> 8$ Myr, star clusters with both canonical and top-heavy IMFs have faded substantially (right panels of Fig.~\ref{fig:clump}) and the cycle could start again if the only feedback mechanism were radiation. If at this time the energy input from SN feedback removes a large fraction of the dwarf galaxy's gas mass from the disc then the individual (faded) YMC remains the only bright source of star light until gas infall can reignite further star formation. Therefore, at any moment throughout this cycle, at most one YMC would be visible in the sample of faint, $z=6$ galaxies and in this case the sizes measured in observations of stellar light from high-redshift low-mass galaxies would be the sizes of individual star formation regions rather than the actual sizes of the host galaxies. Without the almost immediate impact of the YMC on its surroundings, additional massive star clusters could form unaffected by feedback for several Myr.

More massive galaxies ($\log M_{\mathrm{200}} [\Msun] \gtrsim 10.5$) with larger sizes are less influenced by this process. Only the gas clumps closest to the YMC are ionized while clumps at larger distances remain unaffected before SN feedback becomes important. The measured sizes would therefore be more representative of the extent of the gaseous disc. In the size-luminosity relation a transition is expected to occur close to the galaxy mass below which radiation feedback from the current YMC can prevent the formation of additional YMCs within the same galaxy. 

Radiative feedback can therefore explain why typically only one YMC is present in low-mass galaxies at $z=6$, leading to the extremely small ($<40\,\mathrm{pc}$) measured intrinsic sizes as well as the possible change in the slope of the size-luminosity relation in the sample of gravitationally lensed objects from \citet{bouwens_very_2017}. Those authors show that the slope in the luminosity-size relation of their $z=6$ galaxy sample likely changes from $R_{\mathrm{eff}} \propto L^{0.26}$ for bright galaxies ($< -17\,\mathrm{mag}$) to $R_{\mathrm{eff}} \propto L^{0.5}$ for fainter objects, where $R_{\mathrm{eff}}$ is the observed effective radius. Their slope for the bright end matches the extrapolated sizes from blank-field studies, while the fainter objects have inferred sizes that are too small compared to the extrapolated relation. 

In Fig.~\ref{fig:clumpsolar} we showed that the radiative feedback effect is stronger for low-metallicity galaxies, as the additional shielding by dust in high-metallicity galaxies allows clumps with larger Jeans masses to remain neutral. Summarising, the suppression of YMC formation by radiative feedback, as modelled here, is expected to be increasingly efficient for galaxies with smaller sizes, thicker discs, as well as lower metal and dust contents.

As large gas clumps can be fully ionized throughout the galaxy, a high escape fraction of ionizing photons is expected for these extremely compact, high-$z$ galaxies. This is compatible with measurements of individual galaxies with high escape fractions which occur preferentially for starburst galaxies with similar properties as the objects discussed in this work (see discussion in Sec.~\ref{sec:discussion}).

The analytic and numerical approaches presented in this work are idealised. Therefore, we cannot predict the precise mass or size scales below which radiation feedback from a YMC prevents the formation of additional YMCs and produces the observed change in the slope of the luminosity-size relation. However, we do emphasise that such a transition is expected due to the radiative feedback from the YMC.

\section*{Acknowledgements}
The authors thank the referee for constructive suggestions that helped to improve the clarity of the publication.
SP was supported by European Research Council (ERC) Advanced Investigator grant DMIDAS (GA 786910, PI C. S. Frenk). AH thanks the support of the Netherlands Organisation for Scientific Research (NWO) under the VENI project 639.041.644. 



\bibliographystyle{mnras}
\bibliography{bibliography/HighRedshiftYMC} 



\appendix

\section{Photoionization time-scales}\label{sec:timescales}

In the following we show that the time-scale for radiative feedback is almost instantaneous compared to that for SN feedback ($\tau_{\mathrm{SN}}$, see Eq.~\ref{eq:SNtimescale}). 

\noindent
In the absence of recombinations the time to ionize a pure hydrogen cloud $\Delta t$ is the total number of initially neutral hydrogen atoms over the ionizing luminosity in photons per second, $\dot{N}_{\ion{H}{I}}$. For a YMC at the centre of a homogeneous gas cloud with radius $\lambda_J/2$ (with the Jeans length $\lambda_J$ from Eq.~\ref{eq:jeanslength2}) and density $n$ the time is 

\begin{equation}
\begin{aligned}
   \Delta t = 0.0072 \,\mathrm{Myr}\,  &\left ( \frac{\dot{N}_{\ion{H}{I}}}{10^{53}\,\mathrm{s}^{-1}} \right )^{-1} \\
  							& \left( \frac{T}{10^4\,\mathrm{K}} \right )^{3/2} \left (\frac{n}{1 \ccm} \right ) ^{-1/2}  .
\end{aligned}
\end{equation}

\noindent
or 

\begin{equation}
   \Delta t =  \frac{\lambda_{\mathrm{J}}/2}{c} = 0.0019 \,\mathrm{Myr}\, \left( \frac{T}{10^4\,\mathrm{K}} \right )^{1/2} \left (\frac{n}{1 \ccm} \right ) ^{-1/2}  .
\end{equation}

\noindent
if the speed of light $c$ is limiting the ionization at a distance of $\lambda_{\mathrm{J}}/2$. 

\noindent
Since $\Delta t \propto n^{-1/2}$ higher densities can be ionized faster. This case is therefore different than a cloud whose size is independent of the gas density, where more atoms i.e. higher density leads to a larger ionization time and $\Delta t \propto n$. 

For a gas clump (again with radius $\lambda_{\mathrm{J}}/2$) at the edge of the galaxy disc and therefore at a distance $R_{\mathrm{eff}}$, only photons within the solid angle $\Omega = (\lambda_{\mathrm{J}}/2)^2 \pi / R_{\mathrm{eff}}^2$ contribute to the ionization of this gas clump\footnote{This assumes no absorption between the YMC and the gas clump, see Appendix~\ref{sec:absorption} for details.}. The minimum time-scale for this case is: 

\begin{equation}
\begin{aligned}
  \Delta t = 0.014 \,\mathrm{Myr}\,  &\left ( \frac{\dot{N}_{\ion{H}{I}}}{10^{53}\,\mathrm{s}^{-1}} \right )^{-1}   \left ( \frac{M_{\mathrm{200}}}{10^{10}\Msun} \right )^{2/3} \\
  		& \left( \frac{T}{10^4\,\mathrm{K}} \right )^{1/2} \left (\frac{n}{1 \ccm} \right ) ^{1/2}  
 \end{aligned}
\end{equation}

\noindent
or 

\begin{equation}
  \Delta t = \frac{R_{\mathrm{eff}}}{c} = 0.0013 \,\mathrm{Myr}\,   \left ( \frac{M_{\mathrm{200}}}{10^{10}\Msun} \right )^{1/3} 
\end{equation}

\noindent 
for the speed of light limit. 

The time-scale for recombination is defined as
\begin{equation}
	\tau_{\mathrm{rec}} = 1/(\alpha n) = 0.122 \,\mathrm{Myr} \, / n  .
\end{equation}

\noindent
For densities larger than $n_{\mathrm{lim}}$ where $\tau_{\mathrm{rec}}(n_{\mathrm{lim}}) = \Delta t (n_{\mathrm{lim}})$ recombination is relevant for the time evolution of the Str\"omgren sphere. For clouds with a size close to the Jeans length $n_{\mathrm{lim}}$ depends on the ionizing photon flux as well as the gas temperature (via $\lambda_J$):

\begin{equation}
	n_{\mathrm{lim}} = 283 \,\mathrm{cm}^{-3}\, \left( \frac{\dot{N}_{\ion{H}{I}}}{10^{53}\,\mathrm{s}^{-1}} \right)^2 \left (  \frac{T}{10^4 \,\K}  \right )^{-3} \,\mathrm{cm}^{-3}
\end{equation}

\noindent
The time-dependent solution to the Str\"omgren radius, and therefore including recombination, is

\begin{equation}
	r_i = R_S \left ( 1 - e^{-t/\tau_{rec}} \right )  
\end{equation}

\noindent
The time $t_{\mathrm{fS}}$ when the ionization front has reached a radius of $f_S \cdot R_S$ is 

\begin{equation}
	t_{\mathrm{fS}} = - \tau_{\mathrm{rec}} \cdot \ln (1-f_S)
\end{equation}

\noindent
and the relevant ionization time-scale $\tau_{\mathrm{ion}}$ is therefore
 
\begin{equation}
	\tau_{\mathrm{ion}} = 
	\begin{cases} 
    		\Delta t , & \text{if $n<n_{\mathrm{lim}}$}.\\
    		t_{\mathrm{fS}}, & \text{if $n \ge n_{\mathrm{lim}}$}.
	 \end{cases} 
\end{equation}

\noindent
For a smooth transition between the two time-scales we use $f_S = 1 - e^{-1} = 0.63$ so that $t_{\mathrm{fS}} = \tau_{\mathrm{rec}}$.

In Sec.~\ref{sec:homogeneous} the maximum density $n_{\mathrm{max}} (M_{200})$ (Eq.~\ref{eq:nmax}) of a homogeneous medium for which the Str\"omgren radius exceeds the galaxy size, $R_{\mathrm{eff}}$, is derived. The corresponding time-scale is 

\begin{equation}
	\tau_{\mathrm{ion}} = 0.0027 \,\,\mathrm{Myr} \, \left ( \frac{T}{10^4\,\K} \right )^{3/2} \left ( \frac{M_{200}}{10^{10}\,\Msun} \right )^{1/4} 
\end{equation}

\noindent 
where $n_{\mathrm{max}}(M_{200}) < n_{\mathrm{lim}}$ and therefore 

\begin{equation}
	M_{200} > 6.12 \times 10^6 \,\Msun\, \left ( \frac{\dot{N}_{\ion{H}{I}}}{10^{53}\,\mathrm{s}^{-1}}  \right )^{-3} \left ( \frac{T}{10^4\,\K} \right )^6
\end{equation}

\noindent
For the ionizing luminosities from the star cluster models cYMC and tYMC the time-scale $\tau_{\mathrm{ion}}$ is very short relative to both the age of the star cluster and the occurrence of the first SN. This shows that the gas is ionized almost instantaneously compared to $\tau_{\mathrm{SN}}$.

\section{Absorption by lower-density gas}\label{sec:absorption}
The {\textsc{Cloudy}} calculations as well as Eq.~\ref{eq:dclump} assume that the gas between the star cluster and the inner edge of the shell / clump does not absorb any radiation. If the line-of-sight passes through volume filling lower-density gas (with $n_{\mathrm{low}} = f_{\mathrm{low}}\,n$ and  $f_{\mathrm{low}}  < 1$) before it reaches the inner edge of the clump, the ionization depth into the shell can be reduced. If the lower-density gas is neutral, the radiation does not only ionize the particles in the dense shell $N_{\mathrm{shell}} = V_{\mathrm{shell}} \times n$ but also the particles in the lower-density sphere within the shell $N_{\mathrm{low}} = V_{\mathrm{low}} \times  f_{\mathrm{low}} \,n$. 

The effect of including a lower-density medium at distances smaller than $d$ on the thickness of the shell / clump that can be ionized is therefore the ratio between $R_{\mathrm{i}}$ from Eqs.~\ref{eq:Ri} and \ref{eq:Ri2} and $R_{\mathrm{i,low}}$ that takes the low-density gas into account. Analogously to $R_{\mathrm{i}}$, $R_{\mathrm{i,low}}$ is derived by equating the number of atoms in the Str\"omgren sphere to that of the shell plus the lower-density interior:

\begin{eqnarray}
	R_{\mathrm{i,low}}&=& \left [ R_{\mathrm{S}}^3  - d^3 ( f_{\mathrm{low}}  - 1)   \right ]^{1/3} -d \\
	\mathrm{compare:} \quad R_{\mathrm{i}}&=& \left [ R_{\mathrm{S}}^3  + d^3 \right ]^{1/3} -d
\end{eqnarray}

\noindent
For $R_{\mathrm{S}}^3 \gg d^3 ( f_{\mathrm{low}}  - 1)$ the lower density gas does not change the depth into the clump that can be ionized as $R_{\mathrm{i,low}}/R_{\mathrm{i}} \rightarrow 1$, while for $R_{\mathrm{S}}^3 \approx d^3 ( f_{\mathrm{low}}  - 1)$ the ratio $R_{\mathrm{i,low}}/R_{\mathrm{i}} \approx 3.85 [ (2-f_{\mathrm{low}})^{1/3}-1]$. This means that if the sphere is filled with gas that is a factor 2 (10) less dense than the gas in the shell / clump, $R_{\mathrm{i}}$ is reduced by 44\% (8\%). As we assume a clear distinction between the densities of the clump and the surrounding medium, we made the simplifying assumption that the radiation reaches the highest density clump unabsorbed.

\section{HM12 background radiation field}\label{sec:HM12}

In the absence of active star formation, the gas in a galaxy is mainly exposed to the UVB radiation field, which we take from \citetalias{haardt_radiative_2012}.
For a comparison with the results from the YMC radiation fields, the density limit $n_{\mathrm{max}}$ below which a gas clump of size $\lambda_{\mathrm{J}}$ (for $T=10^4\,\K$) can be ionized is also numerically determined for \citetalias{haardt_radiative_2012} at $z=6$. For \textsc{Cloudy} runs with the background radiation field from \citetalias{haardt_radiative_2012}, the external radiation field is defined by its intensity and a plane-parallel geometry is assumed. This is a slightly different setup compared to the simulations for the YMCs, where the strength of the radiation field is defined by a luminosity of the YMC emitting light into a solid angle of $4\pi\,\mathrm{sr}$, and the distance $d$ between the radiating source and the inner edge of the gas clump.

\begin{figure}
	\begin{center}
		\includegraphics[width=\linewidth, trim = 0.2cm 1cm 0.5cm 1.8cm, clip]{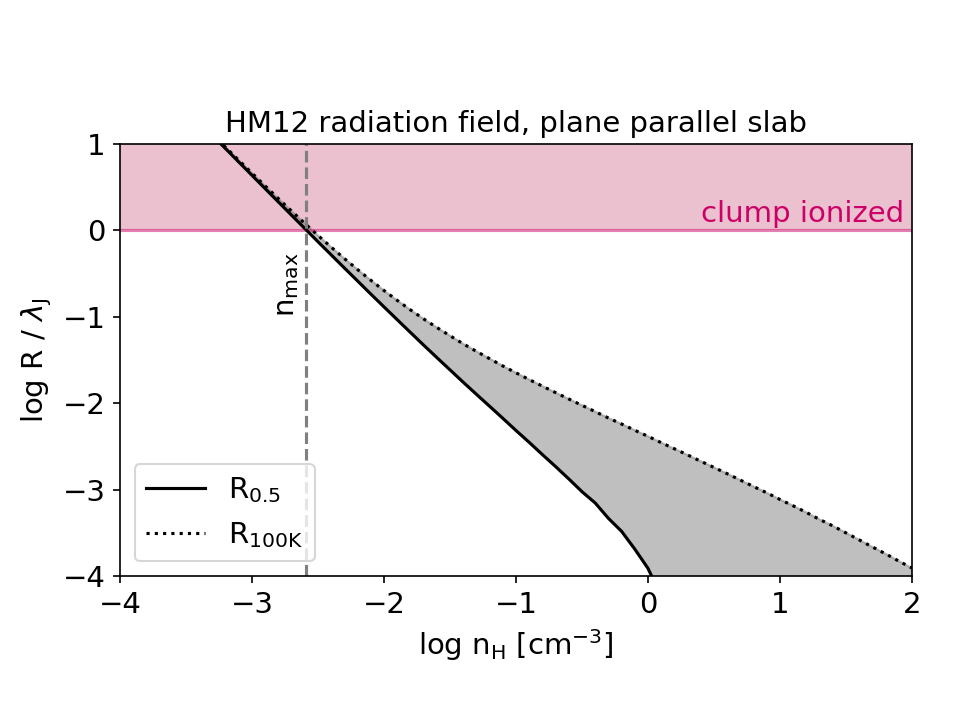}
		\caption{As Fig.~\ref{fig:clumpstoptemp} but for the $z=6$ UVB radiation field from \citetalias{haardt_radiative_2012}. The maximum density of $\log n_{\mathrm{max}} [\ccm] = -2.6$ where gas clumps of size $\lambda_{\mathrm{J}}$ (for $T=10^4\,\K$) can be fully ionized is indicated as vertical dashed line.}
		\label{fig:clumpstoptempHM12}
	\end{center}
\end{figure}

For an individual sightline through the centre of a clump with a size of $\lambda_{\mathrm{J}}$ (Eq.~\ref{eq:jeanslength2}), both $R_{\mathrm{0.5}}/\lambda_{\mathrm{J}}$ and $R_{\mathrm{100K}}/\lambda_{\mathrm{J}}$ are shown in Fig.~\ref{fig:clumpstoptempHM12}. Low-metallicity ($0.1\,\Zsol$), dust-free gas clumps with densities above $\log n_{\mathrm{max}} [\ccm]  = -2.6$ are neutral and cold ($<100\,\K$) if exposed to the  \citetalias{haardt_radiative_2012} UVB radiation field. The maximum density $n_{\mathrm{max}}$ from Fig.~\ref{fig:clumpstoptempHM12} is indicated as reference in Figs.~\ref{fig:clump}~and~\ref{fig:clumpsolar}.

\bsp	
\label{lastpage}
\end{document}